\documentclass[%
 reprint,
superscriptaddress,
 amsmath,amssymb,
 aps,
pra,
]{revtex4-2}
\usepackage{array}
\usepackage{multirow}
\usepackage{graphicx}
\usepackage{subfigure}
\usepackage{bm}
\usepackage{braket}
\usepackage[%
bookmarksnumbered,
bookmarksopen,
colorlinks,
citecolor=blue,
linkcolor=blue,
urlcolor=blue,
]{hyperref}
\begin{document}


\title{Quantum Zeno Effect on Genuine Tripartite Nonlocality and Entanglement in Quantum Dissipative System}


\author{Zi-Yu Xiong}
\affiliation{School of Physics and Electronic Science, Guizhou Normal University, Guiyang 550025, China}
\author{Yong-Jun Xiao}
\affiliation{School of Physics and Electronic Science, Guizhou Normal University, Guiyang 550025, China}
\author{Ye-Qi Zhang}
\affiliation{ Department of Mathematics and Physics, North China Electric Power University, Beijing 102206, China}
\author{Qi-Liang He}\email{heliang005@163.com}
\affiliation{School of Physics and Electronic Science, Guizhou Normal University, Guiyang 550025, China}


\date{\today}

\begin{abstract}
As a precious global resource in quantum information, genuine tripartite nonlocality(GTN) can be quantified by violating Svetlichny inequality. However, there is still no analytical expression for the general three-qubit states due to the difficulty of theoretical calculations. In this paper, we achieve highly accurate quantization of GTN for arbitrary three-qubit quantum states numerically. As an example, we study the dynamics of GTN and genuine tripartite entanglement(GTE) for the W state. Moreover, the complementarity of GTN is verified by examining the nonlocality between the tripartite and the bipartite. Finally, we also find a useful strategy to protect the correlation of GTN and GTE under decoherence by utilizing the Zeno effect.

\end{abstract}\keywords{}


\maketitle

%
\section{introduction}\label{section:1}
Quantum nonlocality, as a significant role in the field of quantum information tasks~\cite{shalm2021device,liu2021device,PhysRevLett.126.050503,10.1063/5.0179566}, was first claimed by Einstein, Podolsky, and Rosen in 1935, which is the well-known EPR paradox~\cite{PhysRev.47.777}.
Then, John Bell established the bell inequality to judge whether the quantum theory satisfied the local realism theory~\cite{bell1964einstein}. The most straightforward Bell-type inequality is the Clauser-Horne-Shimony-Holt (CHSH) inequality for bipartite~\cite{PhysRevLett.23.880}.
If the inequality is violated, the two relevant systems are nonlocal and inseparable, even if they are far apart in space.
Compared to bipartite systems, tripartite systems have significantly more intricate and varied correlation structures~\cite{RevModPhys.86.419}.
Recent experiments have shown that the genuine tripartite nonlocality (GTN) is the strongest three-body correlation and cannot be replicated by any causal theory involving bipartite nonclassical resources~\cite{PhysRevLett.129.060401,hamel2014direct}.
Besides, the relationships between the nonlocality of any subsystem in a tripartite system and the whole have also attracted the attention of many researchers, which are so-called complementarity and monogamy relationships~\cite{PhysRevA.96.022121,toner2009monogamy,PhysRevLett.97.170409}.
The above research demonstrates the unique application value of GTN. (e.g., against a conspiring (cheating) subgroup of parties in the application of quantum communication complexity~\cite{PhysRevA.96.022121,arnon2018practical}.)
Theoretically, Sevtlichny firstly introduced the concept of GTN for tripartite quantum systems and used a Bell-type inequality to quantify GTN in 1987.~\cite{PhysRevD.35.3066}.
Later, based on Sevtlichny's methods, many researchers made important contributions to the detection of GTN.
In 2009, Ghose et al. analyzed the quantitative relationship between tripartite entanglement and GTN for general Greenberger-Horne-Zeilinger (GHZ) pure states~\cite{PhysRevLett.102.250404}. 
Later, Ajoy and Rungta discussed the relationship between GTN for tripartite and the entanglement of its subsystems, which shows that the Svetlichny inequality 
is a more suitable way to measure GTN for W-class states~\cite{PhysRevA.81.052334}.
In 2018, Su et al. derived a method for quantitatively measuring the GTN for general three-qubit states, including pure states and mixed states~\cite{su2018approach}.
This method has been widely adopted by many researchers to effectively measure the GTN of GHZ-class states in the study of relativistic and non-relativistic quantum information~\cite{PhysRevA.107.022201,wu2022genuine,wang2020genuine,wang2020violation}.
However, all the studies mentioned above have been conducted on GHZ-class states. Due to computational complexity, there is no specific analytic expression for detecting GTN for general three-qubit states, which is the main issue that this paper addresses.

Within the framework of quantum correlation, quantum nonlocality arises from quantum entanglement, but quantum entanglement does not necessarily lead to quantum nonlocality~\cite{RevModPhys.86.419,wu2022genuine}.
Like the GTN, the genuine tripartite entanglement (GTE) aims to comprehend the limitations and capabilities in describing the three-way entanglement of complex systems, which has attracted significant attention in both inertial and non-inertial frames~\cite{wu2022genuine,torres2019entanglement,WANG2020135109}.
It is worth mentioning that while there is an inherent connection between GTN and GTE, GTN examines global nonlocal correlations that classical physics can't explain~\cite{PhysRevLett.129.060401}, while GTE focuses on a specific form of entanglement involving all parties~\cite{PhysRevLett.95.090503,PhysRevResearch.4.023059}.
Although many criteria for detecting GTE have been proposed~\cite{PhysRevLett.127.040403,PhysRevResearch.4.023059,Guo_2022,PhysRevA.61.052306}, there are only a few methods to calculate the GTE for mixed states because considering all possible convex roof constructions of a mixed state is difficult in practice.
In general, two methods can be used to quantify GTE for the mixed tripartite state, namely genuinely multipartite concurrence (GMC)~\cite{PhysRevA.86.062303} and $\pi$-tangle~\cite{PhysRevA.75.062308,torres2019entanglement}.
Here, we choose the $\pi$-tangle to compute GTE since the GMC only has analytic expressions for the X-form density matrix. 

Recently, the preparation of W states has received much attention in quantum information. It can be realized experimentally in various quantum systems, such as superconducting quantum system~\cite{kang2016fast,wei2015preparation,neeley2010generation} and atom-cavity coupled systems~\cite{CHEN2016140,Zheng_2005} and so on.
However, real quantum systems inevitably interact with the external environment, leading to decoherence~\cite{schlosshauer2007quantum,10.1063/1.881293} and the rapid disappearance of GTN and GTE. 
This is a significant challenge in achieving quantum information processing.
It is vital to develop effective and practical strategies to protect quantum correlation from the impact of the environment. 
One such strategy is the quantum Zeno effect~\cite{FACCHI1998139,PhysRevLett.100.090503,10.1063/1.523304,PhysRevLett.86.2699}, which involves frequent measurements on the open system to restrict the decay of 
quantum correlations. The quantum Zeno effect has shown promise in protecting entanglement in various scenarios and is 
considered a useful method to combat decoherence~\cite{PhysRevD.29.1626,PhysRevLett.85.1762,PhysRevA.77.062339,PhysRevA.82.052118,PhysRevA.96.032101}. 

In this paper, we numerically achieve the detection of GTN for any tripartite states and verify that our procedure is highly accurate. Based on the procedure for quantifying GTN and $\pi$-tangle,
we obtain the dynamics of GTN and GTE as the function of reservoir correlation time when the tripartite is initially in a max entanglement W state. Our findings reveal that the correlation of GTN and GTE disappears rapidly during the system-environment interaction. 
However, under the effect of environmental memory, GTE will be periodically revived, while GTN will not. In addition, we validate the possibility of utilizing the quantum Zeno effect to protect the system's GTN and GTE. Besides, we also provide a case for the complementarity relationship of GTN by studying the evolution of bell nonlocality between tripartite and its subsystem.

The structure of this paper is as follows:
In Sec.\ref{section:2}, we discuss the definition and calculation methods of GTN and GTE. In addition, the difficulties in calculating the violation of Sevtlichny inequality are described in detail, as well as our solution. 
In Sec.\ref{section:3}, we introduce the model considered and then proceed to study the dynamics of GTN and GTE without the Zeno effect.
In Sec.\ref{section:4}, we briefly introduce the quantum Zeno effect and then investigate the GTN and GTE dynamics under this effect through the same methods consistent with Sec.\ref{section:3}. Finally, a conclusion is given in Sec.\ref{section:5}.
Furthermore, we compared the analytical solution with the numerical solution calculated by our program, and the results were in high agreement with previous studies in Appendix \ref{section:appendix}.
\section{PRELIMINARIES}\label{section:2}
In this section, we first briefly discuss the concepts related to GTN. Then, we elaborate on the challenges associated with computing the maximum violation of Sevtlichny inequality using current theoretical approaches, considering both analytic expressions and numerical simulations, and present our solution. Additionally, we discuss the relevant definition of GTE.
\subsection{Genuine Tripartite Nonlocality}
Generally speaking, a tripartite system's correlation can be termed nonlocal if it cannot be decomposed by the local hidden variable (LHV) model \cite{RevModPhys.86.419,PhysRevLett.47.460,hensen2015loophole,PhysRevLett.118.060401}
\begin{align}
    P(a, b, c \mid x, y, z) & = \sum_{\lambda} q_{\lambda} P_{\lambda}(a \mid x) P_{\lambda}(b \mid y) P_{\lambda}(c \mid z),
\end{align}
where $a, b, c\in\{0,1\}$  represents output for the tripartite Alice, Bob and Charlie sharing the correlation when they receive the input $x, y, z\in\{0,1\}$. $P_{\lambda}(a \mid x)$ is the conditional probability for obtaining 
the output a when the input of Alice is x. However, this criterion is only sufficient to define the nonlocality between two parties since they may share a classical correlation with the third party.
Besides, if two parties share a high degree of nonlocality, they cannot be arbitrarily correlated with any third party, which is the well-known monogamy of nonlocality.  

For GTN, if a correlation of tripartite cannot be decomposed by the following $S_2$ local LHV model, then it can be proven that the correlation is genuine nonlocal~\cite{PhysRevD.35.3066}
\begin{align}
    P(a, b, c \mid x, y, z) & =\sum_{\lambda} q_{\lambda} P_{\lambda}(a \mid x) P_{\lambda}(b, c \mid y, z) \notag \\
    & +\sum_{\mu} q_{\mu} P_{\mu}(b \mid y) P_{\mu}(a, c \mid x, z)\notag \\
    & +\sum_{\nu} q_{\nu} P_{\nu}(c \mid z) P_{\nu}(a, b \mid x, y),
\end{align}
where $\sum_{\lambda} q_{\lambda}+\sum_{\mu} q_{\mu}+\sum_{\nu} q_{\nu}=1$. If a tripartite shares the genuine tripartite nonlocality, there will not exist an arbitrarily
high nonlocal correlation among any subset of the tripartite. More specially, there is no nonlocality between any two parties
of a tripartite system when the tripartite shares an high degree of genuine tripartite nonlocality, which can be called the complementarity of genuine tripartite nonlocality~\cite{PhysRevA.96.022121,toner2009monogamy}.

A practical way to quantitatively analyze the GTN of general three-qubit states is to calculate the maximum violation of the Svetlichny inequality\cite{su2018approach,wang2020violation,su2017generating}
\begin{eqnarray}
    &\operatorname{tr}(S \rho) \leq 4,\\
    S  = & \left(X+X^{\prime}\right) \otimes\left(Y \otimes Z^{\prime}+Y^{\prime} \otimes Z\right)+ \notag \\
    &\left(X-X^{\prime}\right) \otimes\left(Y \otimes Z-Y^{\prime} \otimes Z^{\prime}\right) ,
\end{eqnarray}
considering a general three-qubit system as we discussed earlier, we assume the measurement for Alice are $X=\mathbf{x} \cdot \sigma$ and $X^{\prime}=\mathbf{x}^{\prime} \cdot \sigma$, where
$\mathbf{x}=\left(x_{1}, x_{2}, x_{3}\right)$, $\mathbf{x}^{\prime}=\left(x^{\prime}_{1}, x^{\prime}_{2}, x^{\prime}_{3}\right) \in \mathbb{R}^{3} $ are arbitrary unit vector and $\sigma=\left(\sigma_{1}, \sigma_{2}, \sigma_{3}\right)$ 
is the Pauli matrices. Similarly, we can define the measurement for Bob and Clarlie's system labeled by $\mathbf{b}$, $\mathbf{b}^{\prime}$ and
$\mathbf{c}$, $\mathbf{c}^{\prime}$, respectively.
\begin{align}
    S(\rho_{abc}) \equiv \max _{S} \operatorname{tr}(S \rho_{abc}),
\end{align}
where we define $S(\rho)$ as the maximal value violating the Svetlichny inequality of the tripartite state $\rho_{abc}$ and it can be calculated quantitatively
by the following steps.

The density matrix of any three-qubit can be expressed with the linear combination of Pauli basis
\begin{align}
    \rho=\frac{1}{8} \sum_{i, j, k=0}^{3} t_{i j k} \sigma_{i} \otimes \sigma_{j} \otimes \sigma_{k},
\end{align}
where $t_{i j k}=\operatorname{tr}\left(\rho \sigma_{i} \otimes \sigma_{j} \otimes \sigma_{k}\right)$ and $\sigma_0$ is the identity matrix.
In order to calculate $S (\rho_{abc})$, we need two more vectors
\begin{equation}
    \begin{array}{l}
        \lambda_{0} \equiv T_{z^{\prime}} \mathbf{y}^{T}+T_{z} \mathbf{y}^{T}, \\
        \lambda_{\mathbf{1}} \equiv T_{z} \mathbf{y}^{T}+T_{z^{\prime}} \mathbf{y}^{\prime T},
        \end{array}
\end{equation}
where $T_z=\sum_{k=1}^{3} z_{k} T_{k}$ is the correlation cube, and $T_{k}=\sum_{i=1}^{3}\sum_{j=1}^{3}t_{ijk}$ for $k=1, 2, 3$, which are 3 by 3 real matrices.
These unit vectors $z$, $z^{\prime}$ and $y$, $y^{\prime}$ $\in \mathbb{R}^3$ can be expressed in the following form,
\begin{equation}\label{eq:basis}
\begin{array}{l}
    \left\{\begin{array}{l}
    z_{1}=\sin \alpha_{1} \sin \alpha_{2} \\
    z_{2}=\sin \alpha_{1} \cos \alpha_{2} \\
    z_{3}=\cos \alpha_{1}
    \end{array},\left\{\begin{array}{l}
    z^{\prime}{ }_{1}=\sin \beta_{1} \sin \beta_{2} \\
    z^{\prime}{ }_{2}=\sin \beta_{1} \cos \beta_{2}, \\
    z^{\prime}{ }_{3}=\cos \beta_{1}
    \end{array}\right.\right. \\ \\
    \left\{\begin{array}{l}
    y_{1}=\sin \alpha_{3} \sin \alpha_{4} \\
    y_{2}=\sin \alpha_{3} \cos \alpha_{4} \\
    y_{3}=\cos \alpha_{3}
    \end{array},\left\{\begin{array}{l}
    y_{1}^{\prime}=\sin \beta_{3} \sin \beta_{4} \\
    y_{2}^{\prime}=\sin \beta_{3} \cos \beta_{4} . \\
    y_{3}^{\prime}=\cos \beta_{3}
    \end{array}\right.\right.
    \end{array}
\end{equation} 

The genuine tripartite non-locality of $\rho_{abc}$ can be formulated as 
\begin{align}
    S(\rho_{ABC}) & = 2 \sqrt{F\left(T_{1}, T_{2}, T_{3}\right)},\label{eq:Srho} \\
    F\left(T_{1}, T_{2}, T_{3}\right) & \equiv \max _{\mathbf{y}, \mathbf{y}^{\prime}, \mathbf{z}, \mathbf{z}^{\prime}} \frac{1}{2}
    \bigg[ ( \left\|\lambda_{\mathbf{0}}\right\|^{2}+\left\|\lambda_{\mathbf{1}}\right\|^{2}) \notag  \\ 
    &+\sqrt{(\left\|\lambda_{\mathbf{0}}\right\|^{2}+  \left\|\lambda_{\mathbf{0}}\right\|^{2})^2-4\left\langle\lambda_{0}, \lambda_{\mathbf{1}}\right\rangle^{2}}\bigg],
\end{align}
where $\left\langle\lambda_{0}, \lambda_{\mathbf{1}}\right\rangle^{2}$ defines the inner product between $\lambda_{0}$ and $\lambda_{1}$. 

The method described above can accurately calculate the maximum violation of the Svetlichny inequality, but it is not easy to compute. From the point of view of solving the $S(\rho_{abc})$ expression, theoretical calculations using this method have only been successful in detecting GTN of X-form density matrices for tripartite states, such as the GHZ-class state or some simple mixed states so far~\cite{PhysRevA.107.022201,wu2022genuine,wang2020genuine,wang2020violation,PhysRevA.81.052334}.
Besides, obtaining the analytical expression for $S(\rho_{abc})$ during its dynamic evolution for complex mixed states is difficult because the evolution of the density matrix from $\rho_{(0)}$ to $\rho_{(t)}$ at any moment does not follow simple laws in some cases.
From a numerical perspective, we need to consider all possible Svetlichny operators when calculating the maximum value $S(\rho_{abc})$, equivalent to examining all possible combinations of four sets of unit vectors in the method above. 
However, this process is computationally intensive due to the exponential growth in the number of possible combinations of four groups of bases. Even if numerically considering large angular splits of Eq.~(\ref{eq:basis}), this will result in hundreds of millions of operations for a three-qubit
state. 

However, we accurately calculate $S(\rho_{abc})$ for any three-qubit state by utilizing parallel multi-core allocation on a computer and random sampling in the measurement direction based on the abovementioned calculation methods. 
More specifically, leveraging the multi-core capabilities of the computer workstation, we randomly sample the measurement direction in the remaining three group bases while traversing one of the four bases. The results demonstrate that our numerical program is highly accurate.(See Appendix \ref{section:appendix})

\subsection{Genuine Tripartite Entanglement}
In general, there are only a few ways to effectively calculate the GTE for general three-qubit states since it is diﬀicult to take over all possible decompositions of mixed states. Here, we
choose the method of $\pi$-tangle to quantify the GTE of a three-qubit system, which gives a clear analytical solution, and it can be defined in the following form~\cite{PhysRevA.75.062308,torres2019entanglement}
\begin{align}
    \pi_{ABC}=\frac{1}{3}\left(\pi_{A}+\pi_{B}+\pi_{C}\right),
\end{align}
where, $\pi$-tangle is the average of $\pi_A$, $\pi_B$ and $\pi_C$.
\begin{equation}\label{eq:pi2}
\begin{array}{l}
    \pi_{a}=N_{a(bc)}^{2}-N_{ab}^{2}-N_{ac}^{2}, \\
    \pi_{b}=N_{b(ac)}^{2}-N_{ba}^{2}-N_{bc}^{2}, \\
    \pi_{c}=N_{c(ab)}^{2}-N_{ca}^{2}-N_{cb}^{2},
\end{array}
\end{equation}
where $N_{a(bc)}=\Vert \rho_{abc}^{T_{a}}\Vert-1$ and $N_{ab}=\Vert\rho_{ab}^{T_{a}}\Vert-1$ are the negativity for a tripartite and 
bipartite system, respectively. $T_a$ denotes the partial transpose of $\rho_{abc}$ or $\rho_{ab}$, and $\Vert M \Vert$ is the trace norm for a
matrix M. The other symbols have the same definition like $N_{a(bc)}$ and $N_{ab}$ in Eq.~(\ref{eq:pi2}).

\section{The dynamics of GTN and  GTE without Zeno effect}\label{section:3}
In this section, we consider the case in which three non-interacting qubits interact with a boson environment in equilibrium. The total Hamiltonian of the three-qubit system and boson reservoir is described as~\cite{PhysRevLett.100.090503,Nourmandipour,PhysRevA.79.032310}

\begin{eqnarray}\label{eq:model1}
    H   =  \omega_{0} \sum_{i  = 1}^{3}\left(\sigma_{+}^{(i)}\sigma_{-}^{(i)}\right)+\sum_{k} \omega_{k} a_{k}^{\dagger} a_{k} \notag \\
    +\bigg\{ \alpha_{T} \sum_{i  =  1}^{3} \sum_{k}  g_{k} \sigma_{+}^{(i)} a_{k}+\text { H.c. } \bigg\},
\end{eqnarray}
where $\omega_0$ and $\omega_k$ denote the frequencies of qubit and reservoir.
The spin-flip operators are defined by $\sigma^{(i)}_- = \ket{0_i} \bra{1_i}$, $\sigma^{(i)}_+ = \ket{1_i} \bra{0_i}$
associated with the ground state $\ket{0_i}$ and excited state $\ket{1_i}$ of the $i$th qubit.
Besides, $a_{j}^{\dagger}$ and $a_{j}$ are the creation and annihilation operators of the reservoir and 
$\alpha_T$ is a dimensionless constant
representing the interaction between the qubit and the reservoir.

Suppose that the initial state is set to the W state
\begin{equation}
    |\psi(0)\rangle=\ket{W}\otimes  |0\rangle_{R},
\end{equation}
where $\ket{W}=\frac{1}{\sqrt{3}}[\ket{100}+\ket{010}+\ket{001}  ]$. Here, we consider that the transition frequency $\omega_0$ and constant $\alpha_T$ are same for every qubit. The time evolution of the total system is given by
\begin{equation}\label{eq:model2}
    |\psi(t)\rangle=\mathcal{E}(t)  |W\rangle|0\rangle_{R}+\sum_{k} \Lambda_{k}(t)\ket{G}\ket{1_k}_R ,
\end{equation}
where $\ket{1_k}_R$ is the state of the reservoir with only one excitation in the $k$th mode and $\ket{G}=\ket{000} $.
In the context of the continuum limit for the environment, we are examining a reservoir with a Lorentzian spectral density. This reservoir can be thought of as the electromagnetic field present within a cavity undergoing decoherence. As a result, the spectrum of the field within the cavity can be represented as
\begin{equation}\label{eq:model3}
J(\omega)=\frac{W^{2}}{\pi} \frac{\lambda}{\left(\omega-\omega_{c}\right)^{2}+\lambda^{2}},
\end{equation} 
where $\lambda$ is the width of the distribution with defining the quantity $\frac{1}{\lambda}$ as
the reservoir correlation time. The weight $W$ is proportional to the vacuum Rabi
frequency then $\mathcal{R}=\alpha_{T} W$ is the vacuum Rabi frequency.

By using the Schrödinger equation as well as the Laplace transform with Eqs.~\eqref{eq:model1}--\eqref{eq:model3}, we can get the exact survival probability
$|\mathcal{E}(t)|^{2} \equiv \mathrm{P}_{0}(t)=\left|\left\langle\psi_{0} \mid \psi(t)\right\rangle\right|^{2}$ with
\begin{equation}\label{eq:survival}
    \mathcal{E}(t)=e^{-(\lambda-i \delta) t / 2}\left[\cosh (\Omega t / 2)+\frac{\lambda-i \delta}{\Omega} \sinh (\Omega t / 2)\right],
\end{equation}
where the detuning $\delta=\omega_0 - \omega_k $ and $\Omega =\sqrt{\lambda^{2}-\Omega_{R}^{2}-i 2 \delta \lambda}$ with 
$\Omega_{R}=\sqrt{4 \mathcal{R}^2+\delta^{2}}$.

In the basis $\big\{ |e e e\rangle,|e e g\rangle,|e g e\rangle,|e g g\rangle,|g e e\rangle,|g e g\rangle,|g g e\rangle,$
$|g g g\rangle \big\}$, 
the reduced density matrix for the three qubits is given by
\begin{equation}
    \setlength{\arraycolsep}{2pt}
    \rho_{abc}= \begin{pmatrix}
    0 & 0 & 0 & 0 & 0 & 0 & 0 & 0\\ 
    0 & 0 & 0 & 0 & 0 & 0 & 0 & 0\\
    0 & 0 & 0 & 0 & 0 & 0 & 0 & 0\\[0.5mm]
    0 & 0 & 0 & \frac{\lvert \mathcal{E}(t)  \rvert ^2}{3} & 0  & \frac{\lvert \mathcal{E}(t)  \rvert ^2}{3} & \frac{\lvert \mathcal{E}(t)  \rvert ^2}{3} & 0\\[0.5mm]
    0 & 0 & 0 & 0 & 0 & 0 & 0 & 0\\
    0 & 0 & 0 & \frac{\lvert \mathcal{E}(t)  \rvert ^2}{3} & 0  & \frac{\lvert \mathcal{E}(t)  \rvert ^2}{3} & \frac{\lvert \mathcal{E}(t)  \rvert ^2}{3} & 0\\[0.5mm]
    0 & 0 & 0 & \frac{\lvert \mathcal{E}(t)  \rvert ^2}{3} & 0  & \frac{\lvert \mathcal{E}(t)  \rvert ^2}{3} & \frac{\lvert \mathcal{E}(t)  \rvert ^2}{3} & 0\\[0.5mm]
    0 & 0 & 0 & 0 & 0 & 0 & 0 & 1-\mathcal{E}(t)^2 \\
    \end{pmatrix}.
\end{equation}

Based on the symmetry of the $W$ state, we have $\rho_{ab}=\rho_{ac}=\rho_{bc}$ by tracing over the relevant part of $\rho_{abc}$, then the GTN and GTE can be calculated.

\begin{figure}[htbp]
    \centering
    \includegraphics[scale=0.8]{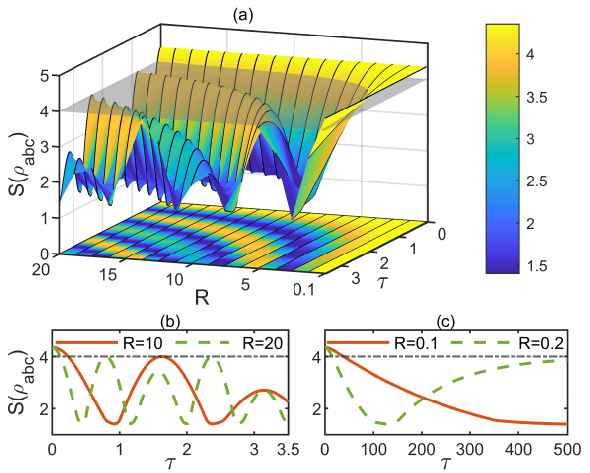}
    \caption{(a)The maximal violation of Sevtlichny inequality $S(\rho_{abc})$ as the function of
    the relative time $\tau$ and the parameter $R=\frac{\mathcal{R}}{\lambda}$. (b)The time evolution of the $S(\rho_{abc})$ in the strong coupling range for
    $R=10$(Solid red line) and $R=20$ (green dotted line). (c)The time evolution of the $S(\rho_{abc})$ in the weak coupling range for
    $R=0.1$(Solid red line) and $R=0.2$ (green dotted line). The $\delta=0$ for all subgraphs.}
    \label{figure:1} 
\end{figure}

Firstly, we look at the evolution of the $S(\rho_{abc})$ when the coupling strength $R=\frac{\lambda}{\mathcal{R}}$ between the three-qubit system and the reservoir changing continuously,
which exhibits different dynamics properties due to memory effect of the reservoir. For $R \gg 1$ and $R \ll 1$, it indicates strong and weak coupling between the qubits and reservoir, respectively.
In Fig.~\ref{figure:1}(a), we plot $S(\rho_{abc})$ as a function of $\tau$ and $R$ and 
the transparent plane of gray for $S(\rho_{abc})=4$. As we can see from the picture, the $S(\rho_{abc})(\tau =0)=4.35$ for $\rho_{abc}(\tau=0)=\ket{W}$ indicates that the
W state shares the correlation of GTN, which is consistent with previous study~\cite{PhysRevA.81.052334}.
As $R$ increases, $S(\rho_{abc})$ decreases more rapidly at first, then periodically revives, but not more than four. At the same time, the frequency of the damping oscillation also increases.
However, the Svetlichny inequality has not been violated, except for the initial time interval. This indicates that we cannot detect the GTN in the subsequent time intervals under the dissipative dynamics of the model.

Next, we will discuss the effect of different coupling strengths on $S(\rho_{abc})$ in more detail.
In Fig.~\ref{figure:1}(b), there exists a time interval at the beginning violating the
Svetlichny inequality and the system shared the GTN. However, the system satisfies Svetlichny inequality in other time intervals as time evolves. During this period, the value of s exhibits a periodic recovery similar to that caused by the memory effect.
Eventually, $S(\rho_{abc})$ will evolve to be close to but less than 4 and become a steady state, such as the green dotted line in Fig.~\ref{figure:1}(c).
In Fig.~\ref{figure:1}(c), the GTN of the three-qubit system can keep GTN for a longer time compared with Fig.~\ref{figure:1}(b), and the $S(\rho_{abc})$
decreases monotonically with $R=0.1$. For $R=0.2$, the $S(\rho_{abc})$ declines more rapidly than $R=0.1$ then stabilizes to a value smaller than 4, reaching a steady state.
The result suggests that the system tends to be steady for a shorter duration as $R$ increases.

As mentioned above, GTN is a global correlation that satisfies complementarity between tripartite and bipartite~\cite{PhysRevA.96.022121}. Here, we detect the nonlocality between the bipartite by using the CHSH inequality~\cite{PhysRevA.101.042112}, and from there we discuss the complementarity relation of the GTN.
\begin{figure}[htbp]
    \centering
    \includegraphics[scale=0.8]{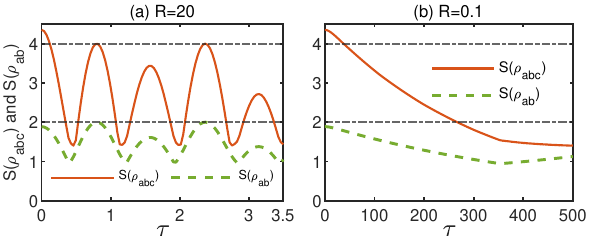}
    \caption{(a)$S(\rho_{abc})$ and $S(\rho_{ab})$ is plotted as a function of $\tau$ for R=20 and R=0.1. }
    \label{figure:2}
\end{figure}

The complementarity of GTN indicates that a three-qubit system exhibiting genuine tripartite nonlocality correlation will impose restrictions on the nonlocality correlation of any subsets of the system. A stronger GTN results in a weaker nonlocality correlation between any two parties and may even eliminate two-way nonlocal correlation altogether.
In Fig.~\ref{figure:2}, for both cases, we find that there is no violation of the CHSH inequality between any two parties of the three-qubit system during the time evolution
since we have $\rho_{ab}$=$\rho_{ac}$=$\rho_{bc}$ due to the symmetry of W state. Additionally, the evolution of $S(\rho_{abc})$ and $S(\rho_{ab})$ is similar but the
$S(\rho_{abc})(\tau=0)=4.35 > 4$ at the beginning, which represents the system share a high degree correlation of GTN.
Due to the complementarity condition, the nonlocality between the two bodies is restricted.

\begin{figure}[htbp]
    \centering
    \includegraphics[scale=0.8]{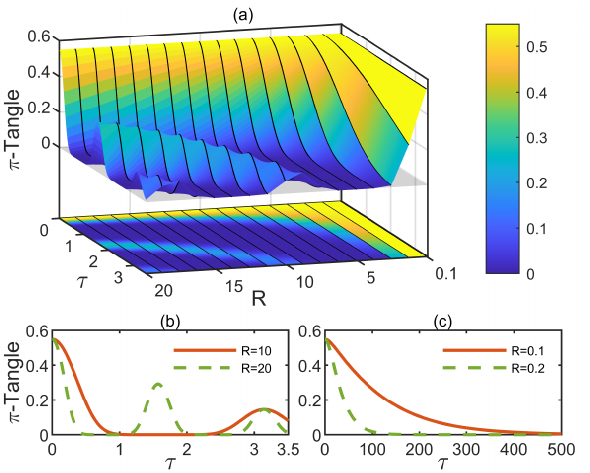}
    \caption{(a) GTE of $\rho_{abc}(t)$ as a function of
    the relative time $\tau$ and of the parameter $R=\frac{\mathcal{R}}{\lambda}$.(b) The time evolution of the GTE in the strong coupling range for
    $R=10$(Solid red line) and $R=20$ (green dotted line). (c) The time evolution of the GTE in the weak coupling range for
    $R=0.1$(Solid red line) and $R=0.2$ (green dotted line). The $\delta=0$ for all subgraphs.}
    \label{figure:3}
\end{figure}
In Fig.~\ref{figure:3}(a), we study the dynamics of GTE versus $\tau=\lambda t$ and $R$,
demonstrating the variation in $\pi$-tangle evolution as their interaction with the environment intensifies.
We observe that the system initially has a GTE correlation, which is then destroyed by decoherence. As the parameter R increases, the rate of GTE decay also increases. However, there is a revival of GTE, indicating that while interactions with the environment can quickly destroy GTE correlation, 
the memory effect of the environment can enable a system lacking global correlation to regain the correlation of GTE, which shows the phenomenon of "sudden death" and "sudden revival" of GTE.
In Fig.~\ref{figure:3}(b) and (c), we compare the dynamics of GTE in the
strong$(R=10,20)$ and weak$(R=0.1,0.2)$ coupling strengths. It shows that under strong coupling, the GTE decays very quickly, and the GTE dies and revives, which presents the phenomenon of periodic damped oscillations of the GTE in response to the memory effect.
When the coupling strength increases, the GTE value
will get a more remarkable recovery, indicating that the memory effect will be more significant as the coupling strength increases.
For weak coupling cases, throughout the evolution, GTE slowly declined and eventually formed a steady state, and the revival phenomenon of GTE has not appeared. 

In the evolution of $S(\rho_{abc})$ and GTE, we did not discuss the case of detuning $\delta \neq 0$ because the effect of detuning is to weaken the interaction between the system and the environment, resulting in the decay of $S(\rho_{abc})$ and GTE slower, which is a trivial example, so we do not discuss it further.

\section{The dynamics of GTN and GTE with Zeno effect}\label{section:4}
In the above, we have confirmed that the $W$ state shares GTN and that this correlation will disappear rapidly since the system-reservoir interaction.
So, it is a natural issue for us to seek a method to protect the GTN and GTE from decoherence.
In this section, we utilize the quantum Zeno effect to effectively protect the correlation of GTN and GTE by conducting a series of frequent measurements.
The measurement of the system can be realized by action of a series of nonselective measurements on the collective atomic (qubits') system, performed at regular time intervals. 
Here, we can detect the energy of all atoms to find whether the three qubits remain in $\left|\psi_{g}\right\rangle=|0\rangle_{1}|0\rangle_{2}|0\rangle_{3}$ state without destroying the entanglement (nonlocality) between qubits~\cite{PhysRevLett.100.090503}.
Take the superconducting qubit system as an example~\cite{Kakuyanagi_2015}, the supercurrent direction is different between the qubit ground and
excited state. By detecting the small magnetic field from the supercurrent, we can distinguish the qubit state.
Besides, we also can test the resonator for the presence of photons since W state is a superposition of three states in each ofwhich one
qubit is excited.
We define $T=\frac{t}{N}$ as the time interval for each measurement, where $N$ is the total number of measurements taken within time $t$.
After performing each measurement, the survival probability with Eq.~(\ref{eq:survival}) will be projected back to its initial state with
relevant probability $P^N (t)=\vert \left\langle\psi(0)| \psi(T)\right\rangle\vert ^{2N}$ and it can be rewritten as~\cite{PhysRevLett.100.090503} 
\begin{equation}
    P^{(N)}(t)=\exp \left[-\Gamma_{z}(T) t\right],
\end{equation}
where, $\Gamma_{z}(T)=-\log \left[|\mathcal{E}(T)|^{2}\right] / T$. 
The relevant quantum correlation can be effectively preserved by frequently resetting the system to its initial state.
\begin{figure}[htbp]
    \centering
    \includegraphics[scale=0.8]{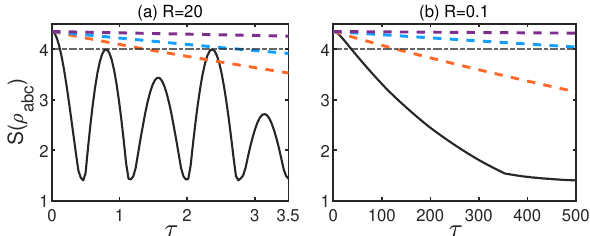}
    \caption{The time evolution of $S(\rho_{abc})$ is plotted as the function of $\tau$ in the absence of measurement(solid black line) and in the presence of measurement performed with different frequencies. (a)The time interval for each measurement is $\lambda t =0.01$, 0.005, 0.001(orange, blue, and purple dotted lines, respectively) with R=20. (b) The time interval for each measurement is $\lambda t =5$, 1, 0.1(orange, blue, and purple dotted line, respectively) with R=0.1.}
    \label{figure:4}
\end{figure}

In this section, we investigate the Zeno effect's impact on the system's GTN and GTE under strong and weak coupling with non-selective measurement.
In Fig.~\ref{figure:4}, the $S(\rho_{abc})$ exhibits a periodic oscillation (R=20) and only shares GTN for a brief period when the system is absent of measurement (solid black line). It shows that the correlation of GTN can be easily destroyed by decoherence. However, by performing frequent, non-selective measurements of the system, we can protect GTN for a long time
as the measurement interval is shortened since the $S(\rho_{abc})$ remains closer to its initial value.
In addition, since the Zeno effect protects the quantum correlation by evolving the quantum state, the GTN's complementarity relationship still holds here.

\begin{figure}[htbp]
    \centering
    \includegraphics[scale=0.8]{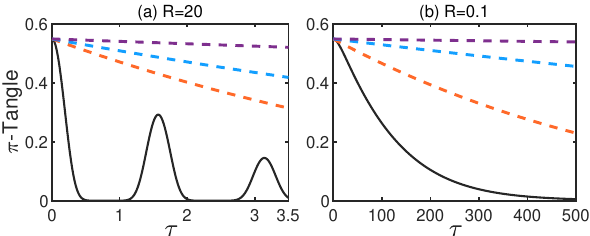}
    \caption{The time evolution of $\pi$-Tangle is plotted as the function of $\tau$ in the absence of measurement(solid black line) and in the presence of measurement performed with different frequencies. (a)The time interval for each measurement is $\lambda t =0.01$, 0.005, 0.001(orange,
    blue, and purple dotted line, respectively) with R=20.
    (b)The time interval for each measurement is $\lambda t =5$, 1, 0.1(orange,
    blue, and purple dotted lines, respectively) with R=0.1.}
    \label{figure:5}
\end{figure}
In Fig.~\ref{figure:5}, we analyze the dynamics of GTE with and without measurements for strong (R=20) and weak (R=0.1) coupled strength cases. We observed that the decay of GTE is noticeably delayed compared to the unmeasured condition(solid black line), and the rate of attenuation is effectively suppressed as the measurement frequency increases for both cases. It is important to note that while the Zeno effect can protect the GTE of the system under decoherence, the decay still occurs faster in the case of strong coupling than in the case of weak coupling.
It is worth mentioning that when the system is detuned, the Zeno effect still dominates the evolution of the whole system in both the Markov$(\delta \leq \lambda)$ and non-Markov$(\delta\leq \mathcal{R})$ regimes.
In addition, we do not discuss the case of large detuning in the paper because the system-environment interaction is very weak at large detuning and exhibits an anti-Zeno effect when the measurement intervals are not short enough, but this is a trivial example and we do not discuss it further.

\section{Summary}\label{section:5}
In this work, we numerically implement a metric of GTN for a generalized three-qubit state using the Svetlichny inequality. Through our numerical program, we predict the evolution of the GTN of a three-qubit coupled to a dissipative cavity, and we also study the evolution of the GTE.  
We have reached the following conclusions: (\romannumeral1) In the exactly solvable model we are considering, the initial state of the three-qubit system is GTN. As the coupling strength increases, the time interval during which the system violates the Svetlichny inequality becomes shorter. Similar to the evolution of GTE, the value of $S(\rho_{abc})$ also periodically revives when the coupling strength is strong enough. However, there is no subsequent violation of the Svetlichny inequality, which indicates that the memory effect of the environment cannot revive the nonlocal correlation of the system.
(\romannumeral2) In the evolution of GTE from weak to strong coupling, unlike GTN, the memory effect can revive the GTE after it is destroyed by decoherence when the coupling strength is strong enough. 
(\romannumeral3) We confirmed the complementarity of GTN, and the results indicate that the high GTN of the system restricts the nonlocality between subsystems. 
(\romannumeral4) We also investigate the quantum Zeno effect on this system. It is worth noting that the GTN and GTE of the system can be effectively protected in a dissipative environment by making use of the quantum Zeno effect, which indicates the feasibility of utilizing frequent measurements to protect quantum correlations.

The discussions in this paper can also be applied to study the GTN of the system in an arbitrary three-qubit model, as well as to the protection of GTN in an open system. Therefore, further investigation using the results obtained in this paper will not only help us understand the properties of GTN, but also provide a theoretical basis for subsequent experimental implementation.

\begin{acknowledgments}
    This work was supported by the National Natural Science Foundation of China under Grant Nos.11364006 and 11805065.
    \end{acknowledgments}
    \appendix
\section{Verification of the accuracy of the numerical calculations}\label{section:appendix} 
Here, we aim to verify the feasibility and correctness of our numerical program by reproducing the results of related papers.

\begin{figure}[htbp]
    \centering
    \includegraphics[scale=0.75]{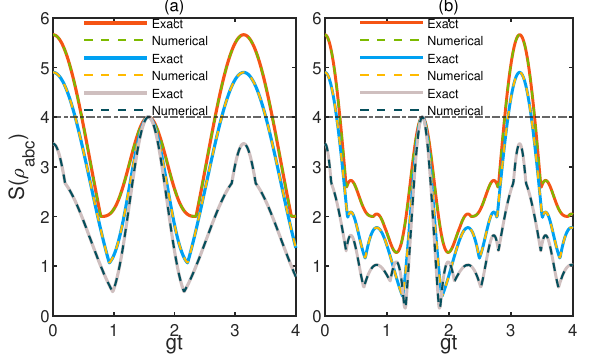}
    \caption{The analytical and numerical result of $\rho_{abc}(t)$ are plotted as a function of
    the time $gt$ for different parameters $\theta=\frac{\pi}{4}$(solid orange line), $\frac{\pi}{6}$(solid blue line), $\frac{\pi}{12}$(solid gray line).
    The three dashed lines are the numerical solutions corresponding to the analytic expression.
    (a) The coupling strengths between the qubits and the cavities are $g_a=g_b=g_c$.
    (b) The coupling strengths between the qubits and the cavities are $g_a=g_b=\frac{g_c}{3}$.}
    \label{figure:appendix1} 
\end{figure}
\begin{figure}[htbp]
    \centering
    \includegraphics[scale=0.75]{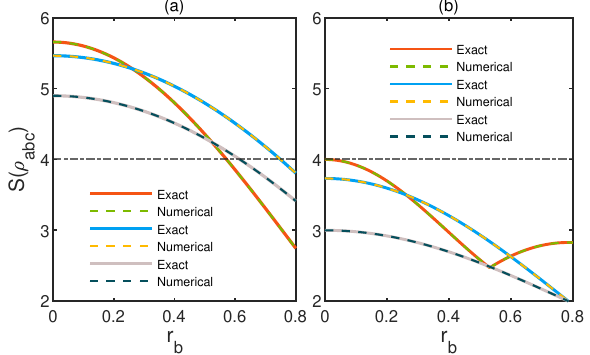}
    \caption{The analytical and numerical results of $\rho_{abc}(t)$ are plotted as the function of
    the acceleration parameters $r_b$ for different acceleration parameters $r_c$.
    The three dashed lines are the numerical results corresponding to the analytic expression.
    (a) $r_c=r_b$(solid orange line) , $r_c=\frac{\pi}{12}$(solid blue line) and $r_c=\frac{\pi}{6}$(solid gray line)
    (b) $r_c=r_b$(solid gray line) , $r_c=\frac{\pi}{12}$(solid blue line) and $r_c=\frac{\pi}{6}$(solid orange line)}
    \label{figure:appendix2} 
\end{figure}

In Fig.~\ref{figure:appendix1}, we calculate two conditions $g_a=g_b=g_c$ or $g_a=g_b=\frac{g_c}{3}$ with different parameters($\theta =\frac{\pi}{4},\frac{\pi}{6},\frac{\pi}{12}$) in the ref~\cite{wang2020violation}, 
and the numerical values are in good agreement with the analytic expression of $S(\rho_{abc})$.
In addition, we calculated some cases in the ref~\cite{wang2020genuine} and the result shows the procedure still can correctly calculate the maximum violation of the Svetlichny inequality in Fig.~\ref{figure:appendix2}.

\begin{table}[!htbp]
    \begin{center}
      \label{tab:table1}
      \begin{tabular}{cp{0.5in}<{\centering}cp{0.7in}<{\centering}c}
      \hline
        Parameters&$S(\rho^{(1)}_{abc})$&Upper Bound of $S(\rho^{(1)}_{abc})$   \\
        \hline
   $p=1,~\theta=\frac{\pi}{3},~\theta_3=\frac{\pi}{2}$ &4.8990&4.8990 \\
  $p=0.8,~\theta=\frac{\pi}{3},~\theta_3=\frac{\pi}{2}$ &3.9192&3.9192 \\
  $p=0.998,~\theta=\frac{\pi}{3},~\theta_3=0.6216$ &3.8610&4.0006 \\
  $p=0.99,~\theta=\frac{\pi}{3},~\theta_3=0.6215$ &3.8298&3.9684 \\
  $p=1,~\theta=\frac{\pi}{4},~\theta_3=\frac{\pi}{2}$ &5.6569&5.6569 \\
  $p=0.8,~\theta=\frac{\pi}{4},~\theta_3=\frac{\pi}{2}$ &4.5255&4.5255 \\
       \hline
      \end{tabular} \label{tab:1} 
    \end{center}
    \caption{The numerical result and theoretical upper bound of the three-qubit GHZ-class states $S(\rho^{(1)}_{abc})$ are calculated under different parameters($p,~\theta,~\theta_3$).}
  \end{table}
  In addition, we compare our work with the existing results. In ref~\cite{PhysRevA.96.042323},  M. Li et.al. proposed a method for calculating the tight upper bound for the maximal quantum value of the Svetlichny operators. 
  \begin{equation}
  \max \left|\langle\mathcal{S}\rangle_{\rho}\right| \leqslant 4 \lambda_{1}
  \end{equation}
  where $\langle\mathcal{S}\rangle_{\rho}=\operatorname{Tr}(\mathcal{S} \rho)$ is the maximal quantum value of the Svetlichny operators. $\lambda_1$ is the maximum singular value of the matrix M (More detials in ref~\cite{PhysRevA.96.042323}).
  In the tabel.~\ref{tab:table1}, we calculate the numerical results (Our Work) and theoretical bounds (M.Li et.al.'s Work) of the maximum violation of the state Svetlichny inequality for the three-qubit GHZ-class states in Eq.~\ref{EQ:A2}. 
  \begin{equation}\label{EQ:A2}
\rho^{(1)}_{abc}=p\left|\psi_{g s}\right\rangle\left\langle\psi_{g s}\right|+\frac{1-p}{8} I,
  \end{equation}
  where $\left|\psi_{g s}\right\rangle=\cos \theta|000\rangle+\sin \theta|11\rangle\left(\cos \theta_{3}|0\rangle+\sin \theta_{3}|1\rangle\right)$
  The results show that our calculations are highly accurate with their criteria. 
  Moreover, when the value of the Upper Bound of $S(\rho^{(1)}_{abc})$ is greater than 4, this only tells us that the system may likely violate the Svetlichny inequality. Nevertheless, our work can give accurate predictions.


\newpage
\bibliography{script.bib}

\begin{thebibliography}{56}%
\makeatletter
\providecommand \@ifxundefined [1]{%
 \@ifx{#1\undefined}
}%
\providecommand \@ifnum [1]{%
 \ifnum #1\expandafter \@firstoftwo
 \else \expandafter \@secondoftwo
 \fi
}%
\providecommand \@ifx [1]{%
 \ifx #1\expandafter \@firstoftwo
 \else \expandafter \@secondoftwo
 \fi
}%
\providecommand \natexlab [1]{#1}%
\providecommand \enquote  [1]{``#1''}%
\providecommand \bibnamefont  [1]{#1}%
\providecommand \bibfnamefont [1]{#1}%
\providecommand \citenamefont [1]{#1}%
\providecommand \href@noop [0]{\@secondoftwo}%
\providecommand \href [0]{\begingroup \@sanitize@url \@href}%
\providecommand \@href[1]{\@@startlink{#1}\@@href}%
\providecommand \@@href[1]{\endgroup#1\@@endlink}%
\providecommand \@sanitize@url [0]{\catcode `\\12\catcode `\$12\catcode
  `\&12\catcode `\#12\catcode `\^12\catcode `\_12\catcode `\%12\relax}%
\providecommand \@@startlink[1]{}%
\providecommand \@@endlink[0]{}%
\providecommand \url  [0]{\begingroup\@sanitize@url \@url }%
\providecommand \@url [1]{\endgroup\@href {#1}{\urlprefix }}%
\providecommand \urlprefix  [0]{URL }%
\providecommand \Eprint [0]{\href }%
\providecommand \doibase [0]{https://doi.org/}%
\providecommand \selectlanguage [0]{\@gobble}%
\providecommand \bibinfo  [0]{\@secondoftwo}%
\providecommand \bibfield  [0]{\@secondoftwo}%
\providecommand \translation [1]{[#1]}%
\providecommand \BibitemOpen [0]{}%
\providecommand \bibitemStop [0]{}%
\providecommand \bibitemNoStop [0]{.\EOS\space}%
\providecommand \EOS [0]{\spacefactor3000\relax}%
\providecommand \BibitemShut  [1]{\csname bibitem#1\endcsname}%
\let\auto@bib@innerbib\@empty
\bibitem [{\citenamefont {Shalm}\ \emph {et~al.}(2021)\citenamefont {Shalm},
  \citenamefont {Zhang}, \citenamefont {Bienfang}, \citenamefont {Schlager},
  \citenamefont {Stevens}, \citenamefont {Mazurek}, \citenamefont
  {Abell{\'a}n}, \citenamefont {Amaya}, \citenamefont {Mitchell}, \citenamefont
  {Alhejji} \emph {et~al.}}]{shalm2021device}%
  \BibitemOpen
  \bibfield  {author} {\bibinfo {author} {\bibfnamefont {L.~K.}\ \bibnamefont
  {Shalm}}, \bibinfo {author} {\bibfnamefont {Y.}~\bibnamefont {Zhang}},
  \bibinfo {author} {\bibfnamefont {J.~C.}\ \bibnamefont {Bienfang}}, \bibinfo
  {author} {\bibfnamefont {C.}~\bibnamefont {Schlager}}, \bibinfo {author}
  {\bibfnamefont {M.~J.}\ \bibnamefont {Stevens}}, \bibinfo {author}
  {\bibfnamefont {M.~D.}\ \bibnamefont {Mazurek}}, \bibinfo {author}
  {\bibfnamefont {C.}~\bibnamefont {Abell{\'a}n}}, \bibinfo {author}
  {\bibfnamefont {W.}~\bibnamefont {Amaya}}, \bibinfo {author} {\bibfnamefont
  {M.~W.}\ \bibnamefont {Mitchell}}, \bibinfo {author} {\bibfnamefont {M.~A.}\
  \bibnamefont {Alhejji}}, \emph {et~al.},\ }\href
  {https://doi.org/10.1038/s41567-020-01153-4} {\bibfield  {journal} {\bibinfo
  {journal} {Nature Physics}\ }\textbf {\bibinfo {volume} {17}},\ \bibinfo
  {pages} {452} (\bibinfo {year} {2021})}\BibitemShut {NoStop}%
\bibitem [{\citenamefont {Liu}\ \emph {et~al.}(2021)\citenamefont {Liu},
  \citenamefont {Li}, \citenamefont {Ragy}, \citenamefont {Zhao}, \citenamefont
  {Bai}, \citenamefont {Liu}, \citenamefont {Brown}, \citenamefont {Zhang},
  \citenamefont {Colbeck}, \citenamefont {Fan} \emph {et~al.}}]{liu2021device}%
  \BibitemOpen
  \bibfield  {author} {\bibinfo {author} {\bibfnamefont {W.-Z.}\ \bibnamefont
  {Liu}}, \bibinfo {author} {\bibfnamefont {M.-H.}\ \bibnamefont {Li}},
  \bibinfo {author} {\bibfnamefont {S.}~\bibnamefont {Ragy}}, \bibinfo {author}
  {\bibfnamefont {S.-R.}\ \bibnamefont {Zhao}}, \bibinfo {author}
  {\bibfnamefont {B.}~\bibnamefont {Bai}}, \bibinfo {author} {\bibfnamefont
  {Y.}~\bibnamefont {Liu}}, \bibinfo {author} {\bibfnamefont {P.~J.}\
  \bibnamefont {Brown}}, \bibinfo {author} {\bibfnamefont {J.}~\bibnamefont
  {Zhang}}, \bibinfo {author} {\bibfnamefont {R.}~\bibnamefont {Colbeck}},
  \bibinfo {author} {\bibfnamefont {J.}~\bibnamefont {Fan}}, \emph {et~al.},\
  }\href {https://doi.org/10.1038/s41567-020-01147-2} {\bibfield  {journal}
  {\bibinfo  {journal} {Nature Physics}\ }\textbf {\bibinfo {volume} {17}},\
  \bibinfo {pages} {448} (\bibinfo {year} {2021})}\BibitemShut {NoStop}%
\bibitem [{\citenamefont {Li}\ \emph {et~al.}(2021)\citenamefont {Li},
  \citenamefont {Zhang}, \citenamefont {Liu}, \citenamefont {Zhao},
  \citenamefont {Bai}, \citenamefont {Liu}, \citenamefont {Zhao}, \citenamefont
  {Peng}, \citenamefont {Zhang}, \citenamefont {Zhang}, \citenamefont {Munro},
  \citenamefont {Ma}, \citenamefont {Zhang}, \citenamefont {Fan},\ and\
  \citenamefont {Pan}}]{PhysRevLett.126.050503}%
  \BibitemOpen
  \bibfield  {author} {\bibinfo {author} {\bibfnamefont {M.-H.}\ \bibnamefont
  {Li}}, \bibinfo {author} {\bibfnamefont {X.}~\bibnamefont {Zhang}}, \bibinfo
  {author} {\bibfnamefont {W.-Z.}\ \bibnamefont {Liu}}, \bibinfo {author}
  {\bibfnamefont {S.-R.}\ \bibnamefont {Zhao}}, \bibinfo {author}
  {\bibfnamefont {B.}~\bibnamefont {Bai}}, \bibinfo {author} {\bibfnamefont
  {Y.}~\bibnamefont {Liu}}, \bibinfo {author} {\bibfnamefont {Q.}~\bibnamefont
  {Zhao}}, \bibinfo {author} {\bibfnamefont {Y.}~\bibnamefont {Peng}}, \bibinfo
  {author} {\bibfnamefont {J.}~\bibnamefont {Zhang}}, \bibinfo {author}
  {\bibfnamefont {Y.}~\bibnamefont {Zhang}}, \bibinfo {author} {\bibfnamefont
  {W.~J.}\ \bibnamefont {Munro}}, \bibinfo {author} {\bibfnamefont
  {X.}~\bibnamefont {Ma}}, \bibinfo {author} {\bibfnamefont {Q.}~\bibnamefont
  {Zhang}}, \bibinfo {author} {\bibfnamefont {J.}~\bibnamefont {Fan}},\ and\
  \bibinfo {author} {\bibfnamefont {J.-W.}\ \bibnamefont {Pan}},\ }\href
  {https://doi.org/10.1103/PhysRevLett.126.050503} {\bibfield  {journal}
  {\bibinfo  {journal} {Phys. Rev. Lett.}\ }\textbf {\bibinfo {volume} {126}},\
  \bibinfo {pages} {050503} (\bibinfo {year} {2021})}\BibitemShut {NoStop}%
\bibitem [{\citenamefont {Zhang}\ \emph {et~al.}(2024)\citenamefont {Zhang},
  \citenamefont {Bian}, \citenamefont {Li}, \citenamefont {Yu},\ and\
  \citenamefont {Guo}}]{10.1063/5.0179566}%
  \BibitemOpen
  \bibfield  {author} {\bibinfo {author} {\bibfnamefont {Y.}~\bibnamefont
  {Zhang}}, \bibinfo {author} {\bibfnamefont {Y.}~\bibnamefont {Bian}},
  \bibinfo {author} {\bibfnamefont {Z.}~\bibnamefont {Li}}, \bibinfo {author}
  {\bibfnamefont {S.}~\bibnamefont {Yu}},\ and\ \bibinfo {author}
  {\bibfnamefont {H.}~\bibnamefont {Guo}},\ }\href
  {https://doi.org/10.1063/5.0179566} {\bibfield  {journal} {\bibinfo
  {journal} {Applied Physics Reviews}\ }\textbf {\bibinfo {volume} {11}},\
  \bibinfo {pages} {011318} (\bibinfo {year} {2024})}\BibitemShut {NoStop}%
\bibitem [{\citenamefont {Einstein}\ \emph {et~al.}(1935)\citenamefont
  {Einstein}, \citenamefont {Podolsky},\ and\ \citenamefont
  {Rosen}}]{PhysRev.47.777}%
  \BibitemOpen
  \bibfield  {author} {\bibinfo {author} {\bibfnamefont {A.}~\bibnamefont
  {Einstein}}, \bibinfo {author} {\bibfnamefont {B.}~\bibnamefont {Podolsky}},\
  and\ \bibinfo {author} {\bibfnamefont {N.}~\bibnamefont {Rosen}},\ }\href
  {https://doi.org/10.1103/PhysRev.47.777} {\bibfield  {journal} {\bibinfo
  {journal} {Phys. Rev.}\ }\textbf {\bibinfo {volume} {47}},\ \bibinfo {pages}
  {777} (\bibinfo {year} {1935})}\BibitemShut {NoStop}%
\bibitem [{\citenamefont {Bell}(1964)}]{bell1964einstein}%
  \BibitemOpen
  \bibfield  {author} {\bibinfo {author} {\bibfnamefont {J.~S.}\ \bibnamefont
  {Bell}},\ }\href@noop {} {\bibfield  {journal} {\bibinfo  {journal} {Physics
  Physique Fizika}\ }\textbf {\bibinfo {volume} {1}},\ \bibinfo {pages} {195}
  (\bibinfo {year} {1964})}\BibitemShut {NoStop}%
\bibitem [{\citenamefont {Clauser}\ \emph {et~al.}(1969)\citenamefont
  {Clauser}, \citenamefont {Horne}, \citenamefont {Shimony},\ and\
  \citenamefont {Holt}}]{PhysRevLett.23.880}%
  \BibitemOpen
  \bibfield  {author} {\bibinfo {author} {\bibfnamefont {J.~F.}\ \bibnamefont
  {Clauser}}, \bibinfo {author} {\bibfnamefont {M.~A.}\ \bibnamefont {Horne}},
  \bibinfo {author} {\bibfnamefont {A.}~\bibnamefont {Shimony}},\ and\ \bibinfo
  {author} {\bibfnamefont {R.~A.}\ \bibnamefont {Holt}},\ }\href
  {https://doi.org/10.1103/PhysRevLett.23.880} {\bibfield  {journal} {\bibinfo
  {journal} {Phys. Rev. Lett.}\ }\textbf {\bibinfo {volume} {23}},\ \bibinfo
  {pages} {880} (\bibinfo {year} {1969})}\BibitemShut {NoStop}%
\bibitem [{\citenamefont {Brunner}\ \emph {et~al.}(2014)\citenamefont
  {Brunner}, \citenamefont {Cavalcanti}, \citenamefont {Pironio}, \citenamefont
  {Scarani},\ and\ \citenamefont {Wehner}}]{RevModPhys.86.419}%
  \BibitemOpen
  \bibfield  {author} {\bibinfo {author} {\bibfnamefont {N.}~\bibnamefont
  {Brunner}}, \bibinfo {author} {\bibfnamefont {D.}~\bibnamefont {Cavalcanti}},
  \bibinfo {author} {\bibfnamefont {S.}~\bibnamefont {Pironio}}, \bibinfo
  {author} {\bibfnamefont {V.}~\bibnamefont {Scarani}},\ and\ \bibinfo {author}
  {\bibfnamefont {S.}~\bibnamefont {Wehner}},\ }\href
  {https://doi.org/10.1103/RevModPhys.86.419} {\bibfield  {journal} {\bibinfo
  {journal} {Rev. Mod. Phys.}\ }\textbf {\bibinfo {volume} {86}},\ \bibinfo
  {pages} {419} (\bibinfo {year} {2014})}\BibitemShut {NoStop}%
\bibitem [{\citenamefont {Huang}\ \emph {et~al.}(2022)\citenamefont {Huang},
  \citenamefont {Gu}, \citenamefont {Jiang}, \citenamefont {Wu}, \citenamefont
  {Bai}, \citenamefont {Chen}, \citenamefont {Sun}, \citenamefont {Zhang},
  \citenamefont {Yu}, \citenamefont {Zhang}, \citenamefont {Lu},\ and\
  \citenamefont {Pan}}]{PhysRevLett.129.060401}%
  \BibitemOpen
  \bibfield  {author} {\bibinfo {author} {\bibfnamefont {L.}~\bibnamefont
  {Huang}}, \bibinfo {author} {\bibfnamefont {X.-M.}\ \bibnamefont {Gu}},
  \bibinfo {author} {\bibfnamefont {Y.-F.}\ \bibnamefont {Jiang}}, \bibinfo
  {author} {\bibfnamefont {D.}~\bibnamefont {Wu}}, \bibinfo {author}
  {\bibfnamefont {B.}~\bibnamefont {Bai}}, \bibinfo {author} {\bibfnamefont
  {M.-C.}\ \bibnamefont {Chen}}, \bibinfo {author} {\bibfnamefont {Q.-C.}\
  \bibnamefont {Sun}}, \bibinfo {author} {\bibfnamefont {J.}~\bibnamefont
  {Zhang}}, \bibinfo {author} {\bibfnamefont {S.}~\bibnamefont {Yu}}, \bibinfo
  {author} {\bibfnamefont {Q.}~\bibnamefont {Zhang}}, \bibinfo {author}
  {\bibfnamefont {C.-Y.}\ \bibnamefont {Lu}},\ and\ \bibinfo {author}
  {\bibfnamefont {J.-W.}\ \bibnamefont {Pan}},\ }\href
  {https://doi.org/10.1103/PhysRevLett.129.060401} {\bibfield  {journal}
  {\bibinfo  {journal} {Phys. Rev. Lett.}\ }\textbf {\bibinfo {volume} {129}},\
  \bibinfo {pages} {060401} (\bibinfo {year} {2022})}\BibitemShut {NoStop}%
\bibitem [{\citenamefont {Hamel}\ \emph {et~al.}(2014)\citenamefont {Hamel},
  \citenamefont {Shalm}, \citenamefont {H{\"u}bel}, \citenamefont {Miller},
  \citenamefont {Marsili}, \citenamefont {Verma}, \citenamefont {Mirin},
  \citenamefont {Nam}, \citenamefont {Resch},\ and\ \citenamefont
  {Jennewein}}]{hamel2014direct}%
  \BibitemOpen
  \bibfield  {author} {\bibinfo {author} {\bibfnamefont {D.~R.}\ \bibnamefont
  {Hamel}}, \bibinfo {author} {\bibfnamefont {L.~K.}\ \bibnamefont {Shalm}},
  \bibinfo {author} {\bibfnamefont {H.}~\bibnamefont {H{\"u}bel}}, \bibinfo
  {author} {\bibfnamefont {A.~J.}\ \bibnamefont {Miller}}, \bibinfo {author}
  {\bibfnamefont {F.}~\bibnamefont {Marsili}}, \bibinfo {author} {\bibfnamefont
  {V.~B.}\ \bibnamefont {Verma}}, \bibinfo {author} {\bibfnamefont {R.~P.}\
  \bibnamefont {Mirin}}, \bibinfo {author} {\bibfnamefont {S.~W.}\ \bibnamefont
  {Nam}}, \bibinfo {author} {\bibfnamefont {K.~J.}\ \bibnamefont {Resch}},\
  and\ \bibinfo {author} {\bibfnamefont {T.}~\bibnamefont {Jennewein}},\ }\href
  {https://doi.org/10.1038/nphoton.2014.218} {\bibfield  {journal} {\bibinfo
  {journal} {Nature Photonics}\ }\textbf {\bibinfo {volume} {8}},\ \bibinfo
  {pages} {801} (\bibinfo {year} {2014})}\BibitemShut {NoStop}%
\bibitem [{\citenamefont {Sami}\ \emph {et~al.}(2017)\citenamefont {Sami},
  \citenamefont {Chakrabarty},\ and\ \citenamefont
  {Chaturvedi}}]{PhysRevA.96.022121}%
  \BibitemOpen
  \bibfield  {author} {\bibinfo {author} {\bibfnamefont {S.}~\bibnamefont
  {Sami}}, \bibinfo {author} {\bibfnamefont {I.}~\bibnamefont {Chakrabarty}},\
  and\ \bibinfo {author} {\bibfnamefont {A.}~\bibnamefont {Chaturvedi}},\
  }\href {https://doi.org/10.1103/PhysRevA.96.022121} {\bibfield  {journal}
  {\bibinfo  {journal} {Phys. Rev. A}\ }\textbf {\bibinfo {volume} {96}},\
  \bibinfo {pages} {022121} (\bibinfo {year} {2017})}\BibitemShut {NoStop}%
\bibitem [{\citenamefont {Toner}(2009)}]{toner2009monogamy}%
  \BibitemOpen
  \bibfield  {author} {\bibinfo {author} {\bibfnamefont {B.}~\bibnamefont
  {Toner}},\ }\href {https://doi.org/10.1098/rspa.2008.0149} {\bibfield
  {journal} {\bibinfo  {journal} {Proceedings of the Royal Society A:
  Mathematical, Physical and Engineering Sciences}\ }\textbf {\bibinfo {volume}
  {465}},\ \bibinfo {pages} {59} (\bibinfo {year} {2009})}\BibitemShut
  {NoStop}%
\bibitem [{\citenamefont {Barrett}\ \emph {et~al.}(2006)\citenamefont
  {Barrett}, \citenamefont {Kent},\ and\ \citenamefont
  {Pironio}}]{PhysRevLett.97.170409}%
  \BibitemOpen
  \bibfield  {author} {\bibinfo {author} {\bibfnamefont {J.}~\bibnamefont
  {Barrett}}, \bibinfo {author} {\bibfnamefont {A.}~\bibnamefont {Kent}},\ and\
  \bibinfo {author} {\bibfnamefont {S.}~\bibnamefont {Pironio}},\ }\href
  {https://doi.org/10.1103/PhysRevLett.97.170409} {\bibfield  {journal}
  {\bibinfo  {journal} {Phys. Rev. Lett.}\ }\textbf {\bibinfo {volume} {97}},\
  \bibinfo {pages} {170409} (\bibinfo {year} {2006})}\BibitemShut {NoStop}%
\bibitem [{\citenamefont {Arnon-Friedman}\ \emph {et~al.}(2018)\citenamefont
  {Arnon-Friedman}, \citenamefont {Dupuis}, \citenamefont {Fawzi},
  \citenamefont {Renner},\ and\ \citenamefont {Vidick}}]{arnon2018practical}%
  \BibitemOpen
  \bibfield  {author} {\bibinfo {author} {\bibfnamefont {R.}~\bibnamefont
  {Arnon-Friedman}}, \bibinfo {author} {\bibfnamefont {F.}~\bibnamefont
  {Dupuis}}, \bibinfo {author} {\bibfnamefont {O.}~\bibnamefont {Fawzi}},
  \bibinfo {author} {\bibfnamefont {R.}~\bibnamefont {Renner}},\ and\ \bibinfo
  {author} {\bibfnamefont {T.}~\bibnamefont {Vidick}},\ }\href
  {https://doi.org/10.1038/s41467-017-02307-4} {\bibfield  {journal} {\bibinfo
  {journal} {Nature communications}\ }\textbf {\bibinfo {volume} {9}},\
  \bibinfo {pages} {459} (\bibinfo {year} {2018})}\BibitemShut {NoStop}%
\bibitem [{\citenamefont {Svetlichny}(1987)}]{PhysRevD.35.3066}%
  \BibitemOpen
  \bibfield  {author} {\bibinfo {author} {\bibfnamefont {G.}~\bibnamefont
  {Svetlichny}},\ }\href {https://doi.org/10.1103/PhysRevD.35.3066} {\bibfield
  {journal} {\bibinfo  {journal} {Phys. Rev. D}\ }\textbf {\bibinfo {volume}
  {35}},\ \bibinfo {pages} {3066} (\bibinfo {year} {1987})}\BibitemShut
  {NoStop}%
\bibitem [{\citenamefont {Ghose}\ \emph {et~al.}(2009)\citenamefont {Ghose},
  \citenamefont {Sinclair}, \citenamefont {Debnath}, \citenamefont {Rungta},\
  and\ \citenamefont {Stock}}]{PhysRevLett.102.250404}%
  \BibitemOpen
  \bibfield  {author} {\bibinfo {author} {\bibfnamefont {S.}~\bibnamefont
  {Ghose}}, \bibinfo {author} {\bibfnamefont {N.}~\bibnamefont {Sinclair}},
  \bibinfo {author} {\bibfnamefont {S.}~\bibnamefont {Debnath}}, \bibinfo
  {author} {\bibfnamefont {P.}~\bibnamefont {Rungta}},\ and\ \bibinfo {author}
  {\bibfnamefont {R.}~\bibnamefont {Stock}},\ }\href
  {https://doi.org/10.1103/PhysRevLett.102.250404} {\bibfield  {journal}
  {\bibinfo  {journal} {Phys. Rev. Lett.}\ }\textbf {\bibinfo {volume} {102}},\
  \bibinfo {pages} {250404} (\bibinfo {year} {2009})}\BibitemShut {NoStop}%
\bibitem [{\citenamefont {Ajoy}\ and\ \citenamefont
  {Rungta}(2010)}]{PhysRevA.81.052334}%
  \BibitemOpen
  \bibfield  {author} {\bibinfo {author} {\bibfnamefont {A.}~\bibnamefont
  {Ajoy}}\ and\ \bibinfo {author} {\bibfnamefont {P.}~\bibnamefont {Rungta}},\
  }\href {https://doi.org/10.1103/PhysRevA.81.052334} {\bibfield  {journal}
  {\bibinfo  {journal} {Phys. Rev. A}\ }\textbf {\bibinfo {volume} {81}},\
  \bibinfo {pages} {052334} (\bibinfo {year} {2010})}\BibitemShut {NoStop}%
\bibitem [{\citenamefont {Su}\ \emph {et~al.}(2018)\citenamefont {Su},
  \citenamefont {Li},\ and\ \citenamefont {Ling}}]{su2018approach}%
  \BibitemOpen
  \bibfield  {author} {\bibinfo {author} {\bibfnamefont {Z.}~\bibnamefont
  {Su}}, \bibinfo {author} {\bibfnamefont {L.}~\bibnamefont {Li}},\ and\
  \bibinfo {author} {\bibfnamefont {J.}~\bibnamefont {Ling}},\ }\href
  {https://doi.org/10.1007/s11128-018-1852-7} {\bibfield  {journal} {\bibinfo
  {journal} {Quantum Information Processing}\ }\textbf {\bibinfo {volume}
  {17}},\ \bibinfo {pages} {1} (\bibinfo {year} {2018})}\BibitemShut {NoStop}%
\bibitem [{\citenamefont {Kumari}\ \emph {et~al.}(2023)\citenamefont {Kumari},
  \citenamefont {Naikoo}, \citenamefont {Ghosh},\ and\ \citenamefont
  {Pan}}]{PhysRevA.107.022201}%
  \BibitemOpen
  \bibfield  {author} {\bibinfo {author} {\bibfnamefont {S.}~\bibnamefont
  {Kumari}}, \bibinfo {author} {\bibfnamefont {J.}~\bibnamefont {Naikoo}},
  \bibinfo {author} {\bibfnamefont {S.}~\bibnamefont {Ghosh}},\ and\ \bibinfo
  {author} {\bibfnamefont {A.~K.}\ \bibnamefont {Pan}},\ }\href
  {https://doi.org/10.1103/PhysRevA.107.022201} {\bibfield  {journal} {\bibinfo
   {journal} {Phys. Rev. A}\ }\textbf {\bibinfo {volume} {107}},\ \bibinfo
  {pages} {022201} (\bibinfo {year} {2023})}\BibitemShut {NoStop}%
\bibitem [{\citenamefont {Wu}\ and\ \citenamefont
  {Zeng}(2022)}]{wu2022genuine}%
  \BibitemOpen
  \bibfield  {author} {\bibinfo {author} {\bibfnamefont {S.-M.}\ \bibnamefont
  {Wu}}\ and\ \bibinfo {author} {\bibfnamefont {H.-S.}\ \bibnamefont {Zeng}},\
  }\href {https://doi.org/10.1140/epjc/s10052-021-09954-4} {\bibfield
  {journal} {\bibinfo  {journal} {The European Physical Journal C}\ }\textbf
  {\bibinfo {volume} {82}},\ \bibinfo {pages} {4} (\bibinfo {year}
  {2022})}\BibitemShut {NoStop}%
\bibitem [{\citenamefont {Wang}\ \emph
  {et~al.}(2020{\natexlab{a}})\citenamefont {Wang}, \citenamefont {Liang},\
  and\ \citenamefont {Zheng}}]{wang2020genuine}%
  \BibitemOpen
  \bibfield  {author} {\bibinfo {author} {\bibfnamefont {K.}~\bibnamefont
  {Wang}}, \bibinfo {author} {\bibfnamefont {Y.}~\bibnamefont {Liang}},\ and\
  \bibinfo {author} {\bibfnamefont {Z.-J.}\ \bibnamefont {Zheng}},\ }\href
  {https://doi.org/10.1007/s11128-020-02645-1} {\bibfield  {journal} {\bibinfo
  {journal} {Quantum Information Processing}\ }\textbf {\bibinfo {volume}
  {19}},\ \bibinfo {pages} {1} (\bibinfo {year}
  {2020}{\natexlab{a}})}\BibitemShut {NoStop}%
\bibitem [{\citenamefont {Wang}\ and\ \citenamefont
  {Zheng}(2020)}]{wang2020violation}%
  \BibitemOpen
  \bibfield  {author} {\bibinfo {author} {\bibfnamefont {K.}~\bibnamefont
  {Wang}}\ and\ \bibinfo {author} {\bibfnamefont {Z.-J.}\ \bibnamefont
  {Zheng}},\ }\href {https://doi.org/10.1038/s41598-020-63236-9} {\bibfield
  {journal} {\bibinfo  {journal} {Scientific Reports}\ }\textbf {\bibinfo
  {volume} {10}},\ \bibinfo {pages} {6621} (\bibinfo {year}
  {2020})}\BibitemShut {NoStop}%
\bibitem [{\citenamefont {Torres-Arenas}\ \emph {et~al.}(2019)\citenamefont
  {Torres-Arenas}, \citenamefont {Dong}, \citenamefont {Sun}, \citenamefont
  {Qiang},\ and\ \citenamefont {Dong}}]{torres2019entanglement}%
  \BibitemOpen
  \bibfield  {author} {\bibinfo {author} {\bibfnamefont {A.~J.}\ \bibnamefont
  {Torres-Arenas}}, \bibinfo {author} {\bibfnamefont {Q.}~\bibnamefont {Dong}},
  \bibinfo {author} {\bibfnamefont {G.-H.}\ \bibnamefont {Sun}}, \bibinfo
  {author} {\bibfnamefont {W.-C.}\ \bibnamefont {Qiang}},\ and\ \bibinfo
  {author} {\bibfnamefont {S.-H.}\ \bibnamefont {Dong}},\ }\href
  {https://doi.org/10.1016/j.physletb.2018.12.010} {\bibfield  {journal}
  {\bibinfo  {journal} {Physics Letters B}\ }\textbf {\bibinfo {volume}
  {789}},\ \bibinfo {pages} {93} (\bibinfo {year} {2019})}\BibitemShut
  {NoStop}%
\bibitem [{\citenamefont {Wang}\ \emph
  {et~al.}(2020{\natexlab{b}})\citenamefont {Wang}, \citenamefont {Wen},
  \citenamefont {Chen},\ and\ \citenamefont {Jing}}]{WANG2020135109}%
  \BibitemOpen
  \bibfield  {author} {\bibinfo {author} {\bibfnamefont {J.}~\bibnamefont
  {Wang}}, \bibinfo {author} {\bibfnamefont {C.}~\bibnamefont {Wen}}, \bibinfo
  {author} {\bibfnamefont {S.}~\bibnamefont {Chen}},\ and\ \bibinfo {author}
  {\bibfnamefont {J.}~\bibnamefont {Jing}},\ }\href
  {https://doi.org/https://doi.org/10.1016/j.physletb.2019.135109} {\bibfield
  {journal} {\bibinfo  {journal} {Physics Letters B}\ }\textbf {\bibinfo
  {volume} {800}},\ \bibinfo {pages} {135109} (\bibinfo {year}
  {2020}{\natexlab{b}})}\BibitemShut {NoStop}%
\bibitem [{\citenamefont {Plenio}(2005)}]{PhysRevLett.95.090503}%
  \BibitemOpen
  \bibfield  {author} {\bibinfo {author} {\bibfnamefont {M.~B.}\ \bibnamefont
  {Plenio}},\ }\href {https://doi.org/10.1103/PhysRevLett.95.090503} {\bibfield
   {journal} {\bibinfo  {journal} {Phys. Rev. Lett.}\ }\textbf {\bibinfo
  {volume} {95}},\ \bibinfo {pages} {090503} (\bibinfo {year}
  {2005})}\BibitemShut {NoStop}%
\bibitem [{\citenamefont {Li}\ and\ \citenamefont
  {Shang}(2022)}]{PhysRevResearch.4.023059}%
  \BibitemOpen
  \bibfield  {author} {\bibinfo {author} {\bibfnamefont {Y.}~\bibnamefont
  {Li}}\ and\ \bibinfo {author} {\bibfnamefont {J.}~\bibnamefont {Shang}},\
  }\href {https://doi.org/10.1103/PhysRevResearch.4.023059} {\bibfield
  {journal} {\bibinfo  {journal} {Phys. Rev. Res.}\ }\textbf {\bibinfo {volume}
  {4}},\ \bibinfo {pages} {023059} (\bibinfo {year} {2022})}\BibitemShut
  {NoStop}%
\bibitem [{\citenamefont {Xie}\ and\ \citenamefont
  {Eberly}(2021)}]{PhysRevLett.127.040403}%
  \BibitemOpen
  \bibfield  {author} {\bibinfo {author} {\bibfnamefont {S.}~\bibnamefont
  {Xie}}\ and\ \bibinfo {author} {\bibfnamefont {J.~H.}\ \bibnamefont
  {Eberly}},\ }\href {https://doi.org/10.1103/PhysRevLett.127.040403}
  {\bibfield  {journal} {\bibinfo  {journal} {Phys. Rev. Lett.}\ }\textbf
  {\bibinfo {volume} {127}},\ \bibinfo {pages} {040403} (\bibinfo {year}
  {2021})}\BibitemShut {NoStop}%
\bibitem [{\citenamefont {Guo}\ \emph {et~al.}(2022)\citenamefont {Guo},
  \citenamefont {Jia}, \citenamefont {Li},\ and\ \citenamefont
  {Huang}}]{Guo_2022}%
  \BibitemOpen
  \bibfield  {author} {\bibinfo {author} {\bibfnamefont {Y.}~\bibnamefont
  {Guo}}, \bibinfo {author} {\bibfnamefont {Y.}~\bibnamefont {Jia}}, \bibinfo
  {author} {\bibfnamefont {X.}~\bibnamefont {Li}},\ and\ \bibinfo {author}
  {\bibfnamefont {L.}~\bibnamefont {Huang}},\ }\href
  {https://doi.org/10.1088/1751-8121/ac5649} {\bibfield  {journal} {\bibinfo
  {journal} {Journal of Physics A: Mathematical and Theoretical}\ }\textbf
  {\bibinfo {volume} {55}},\ \bibinfo {pages} {145303} (\bibinfo {year}
  {2022})}\BibitemShut {NoStop}%
\bibitem [{\citenamefont {Coffman}\ \emph {et~al.}(2000)\citenamefont
  {Coffman}, \citenamefont {Kundu},\ and\ \citenamefont
  {Wootters}}]{PhysRevA.61.052306}%
  \BibitemOpen
  \bibfield  {author} {\bibinfo {author} {\bibfnamefont {V.}~\bibnamefont
  {Coffman}}, \bibinfo {author} {\bibfnamefont {J.}~\bibnamefont {Kundu}},\
  and\ \bibinfo {author} {\bibfnamefont {W.~K.}\ \bibnamefont {Wootters}},\
  }\href {https://doi.org/10.1103/PhysRevA.61.052306} {\bibfield  {journal}
  {\bibinfo  {journal} {Phys. Rev. A}\ }\textbf {\bibinfo {volume} {61}},\
  \bibinfo {pages} {052306} (\bibinfo {year} {2000})}\BibitemShut {NoStop}%
\bibitem [{\citenamefont {Hashemi~Rafsanjani}\ \emph
  {et~al.}(2012)\citenamefont {Hashemi~Rafsanjani}, \citenamefont {Huber},
  \citenamefont {Broadbent},\ and\ \citenamefont
  {Eberly}}]{PhysRevA.86.062303}%
  \BibitemOpen
  \bibfield  {author} {\bibinfo {author} {\bibfnamefont {S.~M.}\ \bibnamefont
  {Hashemi~Rafsanjani}}, \bibinfo {author} {\bibfnamefont {M.}~\bibnamefont
  {Huber}}, \bibinfo {author} {\bibfnamefont {C.~J.}\ \bibnamefont
  {Broadbent}},\ and\ \bibinfo {author} {\bibfnamefont {J.~H.}\ \bibnamefont
  {Eberly}},\ }\href {https://doi.org/10.1103/PhysRevA.86.062303} {\bibfield
  {journal} {\bibinfo  {journal} {Phys. Rev. A}\ }\textbf {\bibinfo {volume}
  {86}},\ \bibinfo {pages} {062303} (\bibinfo {year} {2012})}\BibitemShut
  {NoStop}%
\bibitem [{\citenamefont {Ou}\ and\ \citenamefont
  {Fan}(2007)}]{PhysRevA.75.062308}%
  \BibitemOpen
  \bibfield  {author} {\bibinfo {author} {\bibfnamefont {Y.-C.}\ \bibnamefont
  {Ou}}\ and\ \bibinfo {author} {\bibfnamefont {H.}~\bibnamefont {Fan}},\
  }\href {https://doi.org/10.1103/PhysRevA.75.062308} {\bibfield  {journal}
  {\bibinfo  {journal} {Phys. Rev. A}\ }\textbf {\bibinfo {volume} {75}},\
  \bibinfo {pages} {062308} (\bibinfo {year} {2007})}\BibitemShut {NoStop}%
\bibitem [{\citenamefont {Kang}\ \emph {et~al.}(2016)\citenamefont {Kang},
  \citenamefont {Chen}, \citenamefont {Wu}, \citenamefont {Huang},
  \citenamefont {Song},\ and\ \citenamefont {Xia}}]{kang2016fast}%
  \BibitemOpen
  \bibfield  {author} {\bibinfo {author} {\bibfnamefont {Y.-H.}\ \bibnamefont
  {Kang}}, \bibinfo {author} {\bibfnamefont {Y.-H.}\ \bibnamefont {Chen}},
  \bibinfo {author} {\bibfnamefont {Q.-C.}\ \bibnamefont {Wu}}, \bibinfo
  {author} {\bibfnamefont {B.-H.}\ \bibnamefont {Huang}}, \bibinfo {author}
  {\bibfnamefont {J.}~\bibnamefont {Song}},\ and\ \bibinfo {author}
  {\bibfnamefont {Y.}~\bibnamefont {Xia}},\ }\href
  {https://doi.org/10.1038/srep36737} {\bibfield  {journal} {\bibinfo
  {journal} {Scientific reports}\ }\textbf {\bibinfo {volume} {6}},\ \bibinfo
  {pages} {36737} (\bibinfo {year} {2016})}\BibitemShut {NoStop}%
\bibitem [{\citenamefont {Wei}\ and\ \citenamefont
  {Chen}(2015)}]{wei2015preparation}%
  \BibitemOpen
  \bibfield  {author} {\bibinfo {author} {\bibfnamefont {X.}~\bibnamefont
  {Wei}}\ and\ \bibinfo {author} {\bibfnamefont {M.-F.}\ \bibnamefont {Chen}},\
  }\href {https://doi.org/10.1007/s11128-015-0973-5} {\bibfield  {journal}
  {\bibinfo  {journal} {Quantum Information Processing}\ }\textbf {\bibinfo
  {volume} {14}},\ \bibinfo {pages} {2419} (\bibinfo {year}
  {2015})}\BibitemShut {NoStop}%
\bibitem [{\citenamefont {Neeley}\ \emph {et~al.}(2010)\citenamefont {Neeley},
  \citenamefont {Bialczak}, \citenamefont {Lenander}, \citenamefont {Lucero},
  \citenamefont {Mariantoni}, \citenamefont {O’connell}, \citenamefont
  {Sank}, \citenamefont {Wang}, \citenamefont {Weides}, \citenamefont {Wenner}
  \emph {et~al.}}]{neeley2010generation}%
  \BibitemOpen
  \bibfield  {author} {\bibinfo {author} {\bibfnamefont {M.}~\bibnamefont
  {Neeley}}, \bibinfo {author} {\bibfnamefont {R.~C.}\ \bibnamefont
  {Bialczak}}, \bibinfo {author} {\bibfnamefont {M.}~\bibnamefont {Lenander}},
  \bibinfo {author} {\bibfnamefont {E.}~\bibnamefont {Lucero}}, \bibinfo
  {author} {\bibfnamefont {M.}~\bibnamefont {Mariantoni}}, \bibinfo {author}
  {\bibfnamefont {A.}~\bibnamefont {O’connell}}, \bibinfo {author}
  {\bibfnamefont {D.}~\bibnamefont {Sank}}, \bibinfo {author} {\bibfnamefont
  {H.}~\bibnamefont {Wang}}, \bibinfo {author} {\bibfnamefont {M.}~\bibnamefont
  {Weides}}, \bibinfo {author} {\bibfnamefont {J.}~\bibnamefont {Wenner}},
  \emph {et~al.},\ }\href {https://doi.org/10.1038/nature09418} {\bibfield
  {journal} {\bibinfo  {journal} {Nature}\ }\textbf {\bibinfo {volume} {467}},\
  \bibinfo {pages} {570} (\bibinfo {year} {2010})}\BibitemShut {NoStop}%
\bibitem [{\citenamefont {Chen}\ \emph {et~al.}(2016)\citenamefont {Chen},
  \citenamefont {Huang}, \citenamefont {Song},\ and\ \citenamefont
  {Xia}}]{CHEN2016140}%
  \BibitemOpen
  \bibfield  {author} {\bibinfo {author} {\bibfnamefont {Y.-H.}\ \bibnamefont
  {Chen}}, \bibinfo {author} {\bibfnamefont {B.-H.}\ \bibnamefont {Huang}},
  \bibinfo {author} {\bibfnamefont {J.}~\bibnamefont {Song}},\ and\ \bibinfo
  {author} {\bibfnamefont {Y.}~\bibnamefont {Xia}},\ }\href
  {https://doi.org/https://doi.org/10.1016/j.optcom.2016.05.068} {\bibfield
  {journal} {\bibinfo  {journal} {Optics Communications}\ }\textbf {\bibinfo
  {volume} {380}},\ \bibinfo {pages} {140} (\bibinfo {year}
  {2016})}\BibitemShut {NoStop}%
\bibitem [{\citenamefont {Zheng}(2004)}]{Zheng_2005}%
  \BibitemOpen
  \bibfield  {author} {\bibinfo {author} {\bibfnamefont {S.-B.}\ \bibnamefont
  {Zheng}},\ }\href {https://doi.org/10.1088/1464-4266/7/1/003} {\bibfield
  {journal} {\bibinfo  {journal} {Journal of Optics B: Quantum and
  Semiclassical Optics}\ }\textbf {\bibinfo {volume} {7}},\ \bibinfo {pages}
  {10} (\bibinfo {year} {2004})}\BibitemShut {NoStop}%
\bibitem [{\citenamefont {Schlosshauer}(2007)}]{schlosshauer2007quantum}%
  \BibitemOpen
  \bibfield  {author} {\bibinfo {author} {\bibfnamefont {D.}~\bibnamefont
  {Schlosshauer}},\ }\bibfield  {journal} {\bibinfo  {journal} {The Frontiers
  Collection (Springer-Verlag, 2007)}\ }\href
  {https://doi.org/10.1007/978-3-540-35775-9} {10.1007/978-3-540-35775-9}
  (\bibinfo {year} {2007})\BibitemShut {NoStop}%
\bibitem [{\citenamefont {Zurek}(1991)}]{10.1063/1.881293}%
  \BibitemOpen
  \bibfield  {author} {\bibinfo {author} {\bibfnamefont {W.~H.}\ \bibnamefont
  {Zurek}},\ }\href {https://doi.org/10.1063/1.881293} {\bibfield  {journal}
  {\bibinfo  {journal} {Physics Today}\ }\textbf {\bibinfo {volume} {44}},\
  \bibinfo {pages} {36} (\bibinfo {year} {1991})}\BibitemShut {NoStop}%
\bibitem [{\citenamefont {Facchi}\ and\ \citenamefont
  {Pascazio}(1998)}]{FACCHI1998139}%
  \BibitemOpen
  \bibfield  {author} {\bibinfo {author} {\bibfnamefont {P.}~\bibnamefont
  {Facchi}}\ and\ \bibinfo {author} {\bibfnamefont {S.}~\bibnamefont
  {Pascazio}},\ }\href
  {https://doi.org/https://doi.org/10.1016/S0375-9601(98)00144-3} {\bibfield
  {journal} {\bibinfo  {journal} {Physics Letters A}\ }\textbf {\bibinfo
  {volume} {241}},\ \bibinfo {pages} {139} (\bibinfo {year}
  {1998})}\BibitemShut {NoStop}%
\bibitem [{\citenamefont {Maniscalco}\ \emph {et~al.}(2008)\citenamefont
  {Maniscalco}, \citenamefont {Francica}, \citenamefont {Zaffino},
  \citenamefont {Lo~Gullo},\ and\ \citenamefont
  {Plastina}}]{PhysRevLett.100.090503}%
  \BibitemOpen
  \bibfield  {author} {\bibinfo {author} {\bibfnamefont {S.}~\bibnamefont
  {Maniscalco}}, \bibinfo {author} {\bibfnamefont {F.}~\bibnamefont
  {Francica}}, \bibinfo {author} {\bibfnamefont {R.~L.}\ \bibnamefont
  {Zaffino}}, \bibinfo {author} {\bibfnamefont {N.}~\bibnamefont {Lo~Gullo}},\
  and\ \bibinfo {author} {\bibfnamefont {F.}~\bibnamefont {Plastina}},\ }\href
  {https://doi.org/10.1103/PhysRevLett.100.090503} {\bibfield  {journal}
  {\bibinfo  {journal} {Phys. Rev. Lett.}\ }\textbf {\bibinfo {volume} {100}},\
  \bibinfo {pages} {090503} (\bibinfo {year} {2008})}\BibitemShut {NoStop}%
\bibitem [{\citenamefont {Misra}\ and\ \citenamefont
  {Sudarshan}(1977)}]{10.1063/1.523304}%
  \BibitemOpen
  \bibfield  {author} {\bibinfo {author} {\bibfnamefont {B.}~\bibnamefont
  {Misra}}\ and\ \bibinfo {author} {\bibfnamefont {E.~C.~G.}\ \bibnamefont
  {Sudarshan}},\ }\href {https://doi.org/10.1063/1.523304} {\bibfield
  {journal} {\bibinfo  {journal} {Journal of Mathematical Physics}\ }\textbf
  {\bibinfo {volume} {18}},\ \bibinfo {pages} {756} (\bibinfo {year}
  {1977})}\BibitemShut {NoStop}%
\bibitem [{\citenamefont {Facchi}\ \emph {et~al.}(2001)\citenamefont {Facchi},
  \citenamefont {Nakazato},\ and\ \citenamefont
  {Pascazio}}]{PhysRevLett.86.2699}%
  \BibitemOpen
  \bibfield  {author} {\bibinfo {author} {\bibfnamefont {P.}~\bibnamefont
  {Facchi}}, \bibinfo {author} {\bibfnamefont {H.}~\bibnamefont {Nakazato}},\
  and\ \bibinfo {author} {\bibfnamefont {S.}~\bibnamefont {Pascazio}},\ }\href
  {https://doi.org/10.1103/PhysRevLett.86.2699} {\bibfield  {journal} {\bibinfo
   {journal} {Phys. Rev. Lett.}\ }\textbf {\bibinfo {volume} {86}},\ \bibinfo
  {pages} {2699} (\bibinfo {year} {2001})}\BibitemShut {NoStop}%
\bibitem [{\citenamefont {Joos}(1984)}]{PhysRevD.29.1626}%
  \BibitemOpen
  \bibfield  {author} {\bibinfo {author} {\bibfnamefont {E.}~\bibnamefont
  {Joos}},\ }\href {https://doi.org/10.1103/PhysRevD.29.1626} {\bibfield
  {journal} {\bibinfo  {journal} {Phys. Rev. D}\ }\textbf {\bibinfo {volume}
  {29}},\ \bibinfo {pages} {1626} (\bibinfo {year} {1984})}\BibitemShut
  {NoStop}%
\bibitem [{\citenamefont {Beige}\ \emph {et~al.}(2000)\citenamefont {Beige},
  \citenamefont {Braun}, \citenamefont {Tregenna},\ and\ \citenamefont
  {Knight}}]{PhysRevLett.85.1762}%
  \BibitemOpen
  \bibfield  {author} {\bibinfo {author} {\bibfnamefont {A.}~\bibnamefont
  {Beige}}, \bibinfo {author} {\bibfnamefont {D.}~\bibnamefont {Braun}},
  \bibinfo {author} {\bibfnamefont {B.}~\bibnamefont {Tregenna}},\ and\
  \bibinfo {author} {\bibfnamefont {P.~L.}\ \bibnamefont {Knight}},\ }\href
  {https://doi.org/10.1103/PhysRevLett.85.1762} {\bibfield  {journal} {\bibinfo
   {journal} {Phys. Rev. Lett.}\ }\textbf {\bibinfo {volume} {85}},\ \bibinfo
  {pages} {1762} (\bibinfo {year} {2000})}\BibitemShut {NoStop}%
\bibitem [{\citenamefont {Wang}\ \emph {et~al.}(2008)\citenamefont {Wang},
  \citenamefont {You},\ and\ \citenamefont {Nori}}]{PhysRevA.77.062339}%
  \BibitemOpen
  \bibfield  {author} {\bibinfo {author} {\bibfnamefont {X.-B.}\ \bibnamefont
  {Wang}}, \bibinfo {author} {\bibfnamefont {J.~Q.}\ \bibnamefont {You}},\ and\
  \bibinfo {author} {\bibfnamefont {F.}~\bibnamefont {Nori}},\ }\href
  {https://doi.org/10.1103/PhysRevA.77.062339} {\bibfield  {journal} {\bibinfo
  {journal} {Phys. Rev. A}\ }\textbf {\bibinfo {volume} {77}},\ \bibinfo
  {pages} {062339} (\bibinfo {year} {2008})}\BibitemShut {NoStop}%
\bibitem [{\citenamefont {Francica}\ \emph {et~al.}(2010)\citenamefont
  {Francica}, \citenamefont {Plastina},\ and\ \citenamefont
  {Maniscalco}}]{PhysRevA.82.052118}%
  \BibitemOpen
  \bibfield  {author} {\bibinfo {author} {\bibfnamefont {F.}~\bibnamefont
  {Francica}}, \bibinfo {author} {\bibfnamefont {F.}~\bibnamefont {Plastina}},\
  and\ \bibinfo {author} {\bibfnamefont {S.}~\bibnamefont {Maniscalco}},\
  }\href {https://doi.org/10.1103/PhysRevA.82.052118} {\bibfield  {journal}
  {\bibinfo  {journal} {Phys. Rev. A}\ }\textbf {\bibinfo {volume} {82}},\
  \bibinfo {pages} {052118} (\bibinfo {year} {2010})}\BibitemShut {NoStop}%
\bibitem [{\citenamefont {Zhou}\ \emph {et~al.}(2017)\citenamefont {Zhou},
  \citenamefont {L\"u}, \citenamefont {Zheng},\ and\ \citenamefont
  {Goan}}]{PhysRevA.96.032101}%
  \BibitemOpen
  \bibfield  {author} {\bibinfo {author} {\bibfnamefont {Z.}~\bibnamefont
  {Zhou}}, \bibinfo {author} {\bibfnamefont {Z.}~\bibnamefont {L\"u}}, \bibinfo
  {author} {\bibfnamefont {H.}~\bibnamefont {Zheng}},\ and\ \bibinfo {author}
  {\bibfnamefont {H.-S.}\ \bibnamefont {Goan}},\ }\href
  {https://doi.org/10.1103/PhysRevA.96.032101} {\bibfield  {journal} {\bibinfo
  {journal} {Phys. Rev. A}\ }\textbf {\bibinfo {volume} {96}},\ \bibinfo
  {pages} {032101} (\bibinfo {year} {2017})}\BibitemShut {NoStop}%
\bibitem [{\citenamefont {Aspect}\ \emph {et~al.}(1981)\citenamefont {Aspect},
  \citenamefont {Grangier},\ and\ \citenamefont {Roger}}]{PhysRevLett.47.460}%
  \BibitemOpen
  \bibfield  {author} {\bibinfo {author} {\bibfnamefont {A.}~\bibnamefont
  {Aspect}}, \bibinfo {author} {\bibfnamefont {P.}~\bibnamefont {Grangier}},\
  and\ \bibinfo {author} {\bibfnamefont {G.}~\bibnamefont {Roger}},\ }\href
  {https://doi.org/10.1103/PhysRevLett.47.460} {\bibfield  {journal} {\bibinfo
  {journal} {Phys. Rev. Lett.}\ }\textbf {\bibinfo {volume} {47}},\ \bibinfo
  {pages} {460} (\bibinfo {year} {1981})}\BibitemShut {NoStop}%
\bibitem [{\citenamefont {Hensen}\ \emph {et~al.}(2015)\citenamefont {Hensen},
  \citenamefont {Bernien}, \citenamefont {Dr{\'e}au}, \citenamefont {Reiserer},
  \citenamefont {Kalb}, \citenamefont {Blok}, \citenamefont {Ruitenberg},
  \citenamefont {Vermeulen}, \citenamefont {Schouten}, \citenamefont
  {Abell{\'a}n} \emph {et~al.}}]{hensen2015loophole}%
  \BibitemOpen
  \bibfield  {author} {\bibinfo {author} {\bibfnamefont {B.}~\bibnamefont
  {Hensen}}, \bibinfo {author} {\bibfnamefont {H.}~\bibnamefont {Bernien}},
  \bibinfo {author} {\bibfnamefont {A.~E.}\ \bibnamefont {Dr{\'e}au}}, \bibinfo
  {author} {\bibfnamefont {A.}~\bibnamefont {Reiserer}}, \bibinfo {author}
  {\bibfnamefont {N.}~\bibnamefont {Kalb}}, \bibinfo {author} {\bibfnamefont
  {M.~S.}\ \bibnamefont {Blok}}, \bibinfo {author} {\bibfnamefont
  {J.}~\bibnamefont {Ruitenberg}}, \bibinfo {author} {\bibfnamefont {R.~F.}\
  \bibnamefont {Vermeulen}}, \bibinfo {author} {\bibfnamefont {R.~N.}\
  \bibnamefont {Schouten}}, \bibinfo {author} {\bibfnamefont {C.}~\bibnamefont
  {Abell{\'a}n}}, \emph {et~al.},\ }\href {https://doi.org/10.1038/nature15759}
  {\bibfield  {journal} {\bibinfo  {journal} {Nature}\ }\textbf {\bibinfo
  {volume} {526}},\ \bibinfo {pages} {682} (\bibinfo {year}
  {2015})}\BibitemShut {NoStop}%
\bibitem [{\citenamefont {Handsteiner}\ \emph {et~al.}(2017)\citenamefont
  {Handsteiner}, \citenamefont {Friedman}, \citenamefont {Rauch}, \citenamefont
  {Gallicchio}, \citenamefont {Liu}, \citenamefont {Hosp}, \citenamefont
  {Kofler}, \citenamefont {Bricher}, \citenamefont {Fink}, \citenamefont
  {Leung}, \citenamefont {Mark}, \citenamefont {Nguyen}, \citenamefont
  {Sanders}, \citenamefont {Steinlechner}, \citenamefont {Ursin}, \citenamefont
  {Wengerowsky}, \citenamefont {Guth}, \citenamefont {Kaiser}, \citenamefont
  {Scheidl},\ and\ \citenamefont {Zeilinger}}]{PhysRevLett.118.060401}%
  \BibitemOpen
  \bibfield  {author} {\bibinfo {author} {\bibfnamefont {J.}~\bibnamefont
  {Handsteiner}}, \bibinfo {author} {\bibfnamefont {A.~S.}\ \bibnamefont
  {Friedman}}, \bibinfo {author} {\bibfnamefont {D.}~\bibnamefont {Rauch}},
  \bibinfo {author} {\bibfnamefont {J.}~\bibnamefont {Gallicchio}}, \bibinfo
  {author} {\bibfnamefont {B.}~\bibnamefont {Liu}}, \bibinfo {author}
  {\bibfnamefont {H.}~\bibnamefont {Hosp}}, \bibinfo {author} {\bibfnamefont
  {J.}~\bibnamefont {Kofler}}, \bibinfo {author} {\bibfnamefont
  {D.}~\bibnamefont {Bricher}}, \bibinfo {author} {\bibfnamefont
  {M.}~\bibnamefont {Fink}}, \bibinfo {author} {\bibfnamefont {C.}~\bibnamefont
  {Leung}}, \bibinfo {author} {\bibfnamefont {A.}~\bibnamefont {Mark}},
  \bibinfo {author} {\bibfnamefont {H.~T.}\ \bibnamefont {Nguyen}}, \bibinfo
  {author} {\bibfnamefont {I.}~\bibnamefont {Sanders}}, \bibinfo {author}
  {\bibfnamefont {F.}~\bibnamefont {Steinlechner}}, \bibinfo {author}
  {\bibfnamefont {R.}~\bibnamefont {Ursin}}, \bibinfo {author} {\bibfnamefont
  {S.}~\bibnamefont {Wengerowsky}}, \bibinfo {author} {\bibfnamefont {A.~H.}\
  \bibnamefont {Guth}}, \bibinfo {author} {\bibfnamefont {D.~I.}\ \bibnamefont
  {Kaiser}}, \bibinfo {author} {\bibfnamefont {T.}~\bibnamefont {Scheidl}},\
  and\ \bibinfo {author} {\bibfnamefont {A.}~\bibnamefont {Zeilinger}},\ }\href
  {https://doi.org/10.1103/PhysRevLett.118.060401} {\bibfield  {journal}
  {\bibinfo  {journal} {Phys. Rev. Lett.}\ }\textbf {\bibinfo {volume} {118}},\
  \bibinfo {pages} {060401} (\bibinfo {year} {2017})}\BibitemShut {NoStop}%
\bibitem [{\citenamefont {Su}(2017)}]{su2017generating}%
  \BibitemOpen
  \bibfield  {author} {\bibinfo {author} {\bibfnamefont {Z.}~\bibnamefont
  {Su}},\ }\href {https://doi.org/10.1007/s11128-016-1493-7} {\bibfield
  {journal} {\bibinfo  {journal} {Quantum Information Processing}\ }\textbf
  {\bibinfo {volume} {16}},\ \bibinfo {pages} {28} (\bibinfo {year}
  {2017})}\BibitemShut {NoStop}%
\bibitem [{\citenamefont {Nourmandipour}\ \emph {et~al.}(2016)\citenamefont
  {Nourmandipour}, \citenamefont {Tavassoly},\ and\ \citenamefont
  {Bolorizadeh}}]{Nourmandipour}%
  \BibitemOpen
  \bibfield  {author} {\bibinfo {author} {\bibfnamefont {A.}~\bibnamefont
  {Nourmandipour}}, \bibinfo {author} {\bibfnamefont {M.~K.}\ \bibnamefont
  {Tavassoly}},\ and\ \bibinfo {author} {\bibfnamefont {M.~A.}\ \bibnamefont
  {Bolorizadeh}},\ }\href {https://doi.org/10.1364/JOSAB.33.001723} {\bibfield
  {journal} {\bibinfo  {journal} {J. Opt. Soc. Am. B}\ }\textbf {\bibinfo
  {volume} {33}},\ \bibinfo {pages} {1723} (\bibinfo {year}
  {2016})}\BibitemShut {NoStop}%
\bibitem [{\citenamefont {Francica}\ \emph {et~al.}(2009)\citenamefont
  {Francica}, \citenamefont {Maniscalco}, \citenamefont {Piilo}, \citenamefont
  {Plastina},\ and\ \citenamefont {Suominen}}]{PhysRevA.79.032310}%
  \BibitemOpen
  \bibfield  {author} {\bibinfo {author} {\bibfnamefont {F.}~\bibnamefont
  {Francica}}, \bibinfo {author} {\bibfnamefont {S.}~\bibnamefont
  {Maniscalco}}, \bibinfo {author} {\bibfnamefont {J.}~\bibnamefont {Piilo}},
  \bibinfo {author} {\bibfnamefont {F.}~\bibnamefont {Plastina}},\ and\
  \bibinfo {author} {\bibfnamefont {K.-A.}\ \bibnamefont {Suominen}},\ }\href
  {https://doi.org/10.1103/PhysRevA.79.032310} {\bibfield  {journal} {\bibinfo
  {journal} {Phys. Rev. A}\ }\textbf {\bibinfo {volume} {79}},\ \bibinfo
  {pages} {032310} (\bibinfo {year} {2009})}\BibitemShut {NoStop}%
\bibitem [{\citenamefont {Su}\ \emph {et~al.}(2020)\citenamefont {Su},
  \citenamefont {Tan},\ and\ \citenamefont {Li}}]{PhysRevA.101.042112}%
  \BibitemOpen
  \bibfield  {author} {\bibinfo {author} {\bibfnamefont {Z.}~\bibnamefont
  {Su}}, \bibinfo {author} {\bibfnamefont {H.}~\bibnamefont {Tan}},\ and\
  \bibinfo {author} {\bibfnamefont {X.}~\bibnamefont {Li}},\ }\href
  {https://doi.org/10.1103/PhysRevA.101.042112} {\bibfield  {journal} {\bibinfo
   {journal} {Phys. Rev. A}\ }\textbf {\bibinfo {volume} {101}},\ \bibinfo
  {pages} {042112} (\bibinfo {year} {2020})}\BibitemShut {NoStop}%
\bibitem [{\citenamefont {Kakuyanagi}\ \emph {et~al.}(2015)\citenamefont
  {Kakuyanagi}, \citenamefont {Baba}, \citenamefont {Matsuzaki}, \citenamefont
  {Nakano}, \citenamefont {Saito},\ and\ \citenamefont
  {Semba}}]{Kakuyanagi_2015}%
  \BibitemOpen
  \bibfield  {author} {\bibinfo {author} {\bibfnamefont {K.}~\bibnamefont
  {Kakuyanagi}}, \bibinfo {author} {\bibfnamefont {T.}~\bibnamefont {Baba}},
  \bibinfo {author} {\bibfnamefont {Y.}~\bibnamefont {Matsuzaki}}, \bibinfo
  {author} {\bibfnamefont {H.}~\bibnamefont {Nakano}}, \bibinfo {author}
  {\bibfnamefont {S.}~\bibnamefont {Saito}},\ and\ \bibinfo {author}
  {\bibfnamefont {K.}~\bibnamefont {Semba}},\ }\href
  {https://doi.org/10.1088/1367-2630/17/6/063035} {\bibfield  {journal}
  {\bibinfo  {journal} {New Journal of Physics}\ }\textbf {\bibinfo {volume}
  {17}},\ \bibinfo {pages} {063035} (\bibinfo {year} {2015})}\BibitemShut
  {NoStop}%
\bibitem [{\citenamefont {Li}\ \emph {et~al.}(2017)\citenamefont {Li},
  \citenamefont {Shen}, \citenamefont {Jing}, \citenamefont {Fei},\ and\
  \citenamefont {Li-Jost}}]{PhysRevA.96.042323}%
  \BibitemOpen
  \bibfield  {author} {\bibinfo {author} {\bibfnamefont {M.}~\bibnamefont
  {Li}}, \bibinfo {author} {\bibfnamefont {S.}~\bibnamefont {Shen}}, \bibinfo
  {author} {\bibfnamefont {N.}~\bibnamefont {Jing}}, \bibinfo {author}
  {\bibfnamefont {S.-M.}\ \bibnamefont {Fei}},\ and\ \bibinfo {author}
  {\bibfnamefont {X.}~\bibnamefont {Li-Jost}},\ }\href
  {https://doi.org/10.1103/PhysRevA.96.042323} {\bibfield  {journal} {\bibinfo
  {journal} {Phys. Rev. A}\ }\textbf {\bibinfo {volume} {96}},\ \bibinfo
  {pages} {042323} (\bibinfo {year} {2017})}\BibitemShut {NoStop}%
\end{thebibliography}%
\end{document}